\documentclass{interact}
\usepackage{graphicx}
\usepackage{comment}
\usepackage{amssymb, amsfonts, amsmath}
\usepackage{braket}
\usepackage{bm}
\usepackage{wrapfig}
\usepackage{hyperref}
\usepackage[T1]{fontenc}
\usepackage[backend=biber,sorting=none,style=numeric-comp]{biblatex}
\usepackage{xcolor}
\usepackage{physics}
\usepackage[normalem]{ulem}
\usepackage{tikz}
\usetikzlibrary{arrows.meta}
\usetikzlibrary{external}

\newcommand{\CCBYFour}{%
  \href{https://creativecommons.org/licenses/by/4.0/}{CC BY 4.0}%
}
\newcommand{\up}{\uparrow}
\newcommand{\down}{\downarrow}
\title{Neural quantum states in condensed matter: advances, best practices, and prospects}

\author{
\name{Jonas B.~Rigo\textsuperscript{1}, Björn J. Wurst\textsuperscript{3}, Rajah Nutakki\textsuperscript{4}, Markus Schmitt\textsuperscript{1,5}, Dante Kennes \textsuperscript{2,3}}
\affil{\textsuperscript{1}Institute of Theoretical Physics, University of Regensburg, 93053 Regensburg, Germany}
\affil{\textsuperscript{2}Institut f\"ur Theorie der Statistischen Physik, RWTH Aachen, 52056 Aachen, Germany and JARA - Fundamentals of Future Information Technology}
\affil{\textsuperscript{3}Max Planck Institute for the Structure and Dynamics of Matter,
Center for Free-Electron Laser Science (CFEL),
Luruper Chaussee 149, 22761 Hamburg, Germany}
\affil{\textsuperscript{4}École Polytechnique}
\affil{\textsuperscript{5}Forschungszentrum Jülich GmbH, Institute of Quantum Control, Peter Grünberg Institut (PGI-8), 52425 Jülich, Germany}
}

\date{\today}
\begin{document}
\maketitle

\definecolor{bgblue}{RGB}{214,241,248}
\definecolor{bgorange}{RGB}{252,228,198}
\definecolor{tnblue}{RGB}{30,110,160}
\definecolor{nqsred}{RGB}{160,35,30}
\definecolor{dmftorange}{RGB}{235,140,20}

\section{Introduction}

The investigation of emergent phenomena in condensed matter occupies a distinctive position in modern physics: the microscopic laws governing electrons and nuclei are known, yet describing their collective consequences remains notoriously difficult~\cite{sommerfeldZurQuantentheorieSpektrallinien1916,MoreAdersonDifferent}. In practice, one therefore works with low-energy effective Hamiltonians or reduced descriptions that retain only the degrees of freedom relevant to the phenomena of interest~\cite{metznerFunctionalRenormalizationGroup2012,bornZurQuantentheorieMolekeln1927a,marzariMaximallyLocalizedWannier2012,changDownfoldingInitioInteracting2024,aryasetiawanFrequencydependentLocalInteractions2004a}. Even after this reduction, most lattice models remain non-integrable, and exact solutions are rapidly ruled out by the exponential growth of the Hilbert-space dimension with system size~\cite{lauchliGroundstateEnergySpin2011,weisseExactDiagonalizationTechniques2008,nandyQuantumDynamicsKrylov2025}. Analytical progress is possible in special limits, through low-dimensional effective theories, field-theoretic mappings, and integrable constructions~\cite{jamesNonperturbativeMethodologiesLowdimensional2018,thackerExactIntegrabilityQuantum1981}, but the generic strongly correlated problem must ultimately be addressed numerically.

\begin{figure}[t!]
    \centering
    \includegraphics[width=0.85\columnwidth]{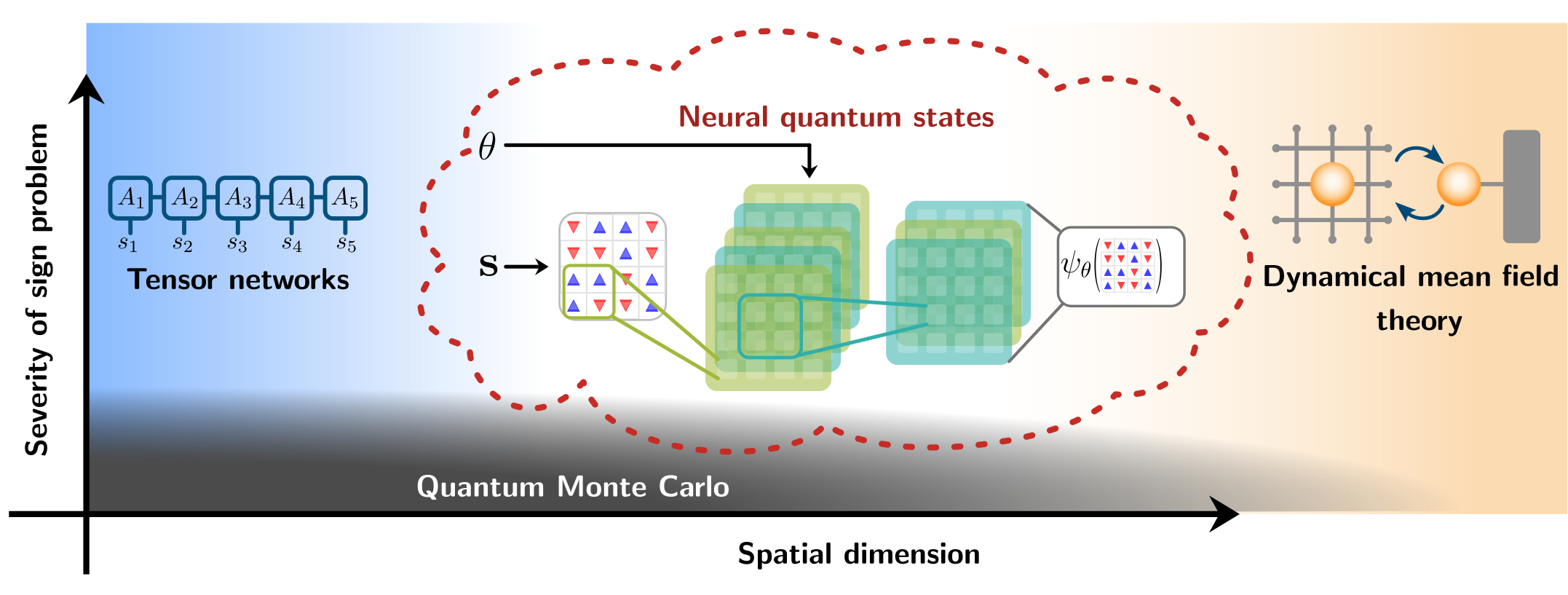}
    \caption{The figure portrays the schematic landscape of the main computational approaches to strongly correlated quantum many-body systems. The vertical axis indicates the severity of the sign problem, while the horizontal axis shows spatial dimension, used here as a practical proxy for the growing entanglement complexity that limits wave-function-based methods. Tensor-network methods are most effective in low dimensions, quantum Monte Carlo performs best when the sign problem is weak or absent, and DMFT is accurate in the high-dimensional limit. Neural quantum states (NQS) are shown as targeting the broad intermediate regime between these established approaches. The illustrated NQS is a \textit{convolutional neural network} on a $4\times4$ spin lattice.}
    \label{fig:sign_vs_dim}
\end{figure}

From a broad perspective, classical simulations of quantum many-body systems can be arranged as in Fig.~\ref{fig:sign_vs_dim}, according to two dominant obstacles:
severity of the sign problem and spatial dimension.
The dimension-related difficulty is most prominent in wave-function-based approaches.
These approaches represent the quantum state explicitly or in compressed form, and include exact diagonalization and Krylov methods~\cite{weisseExactDiagonalizationTechniques2008,nandyQuantumDynamicsKrylov2025}, configuration interaction~\cite{szabo2012modern}, coupled-cluster methods~\cite{szabo2012modern,coupledcluster1}, and tensor-network states~\cite{Schollw_ck_2011,Orus2019}.
Although important exceptions exist~\cite{schindlerVariationalAnsatzGround2022}, a common limitation among wave-function-based methods is entanglement: as entanglement grows with system size, dimension, or time evolution, the cost of representing the state accurately rises rapidly, making generic two- and three-dimensional problems difficult.

Stochastic approaches, most notably quantum Monte Carlo \cite{assaadEvertz2008,becca2017}, avoid representing the full wave function explicitly and instead estimate observables by statistically accumulating contributions from sampled configurations. Their main limitation is the \emph{sign problem}, which can make the variance grow exponentially and thereby render the simulation impractical in precisely the physically interesting regimes~\cite{troyerComputationalComplexityFundamental2005}.
This same stochastic machinery also underlies many of the impurity solvers used in dynamical mean-field theory (DMFT), whose central approximation maps the lattice problem onto a self-consistent quantum impurity model~\cite{georges1996dynamical,maier2005quantum,toschi2007dynamical,rohringer2018diagrammatic}. DMFT becomes exact in the infinite-dimensional limit, but in the physically important regime of intermediate dimensions (2D and 3D), quantum Monte Carlo is hindered by the sign problem while DMFT and its extensions cede the advantage of the underlying infinite-dimensional approximation.

Thus, the intermediate regime remains unsettled in the sense that neither tensor-network methods nor conventional quantum Monte Carlo are uniformly satisfactory there. Many central condensed-matter problems are believed to fall into this regime, including frustrated magnets, doped fermionic lattice models, and non-equilibrium dynamics beyond one dimension~\cite{balents2010,leblanc2015,schmitt2020quantum}.
Neural quantum states---which we discuss in the following---constitute an emerging numerical technique aimed at filling this gap in the computational toolbox by introducing artificial neural network representations of the quantum wave function~\cite{carleoSolvingQuantumManybody2017b}.

The central idea of NQS is to combine a flexible parametrization of the many-body wave function with stochastic evaluation of observables.
Artificial neural networks are used for dimensional reduction to achieve a compressed representation of the wave function.
Monte Carlo sampling of computational states makes it possible to estimate observables without summing over the exponentially large Hilbert space.
This approach does not exhibit the conventional Monte Carlo sign problem~\cite{Nest2009SimulatingQC}; however, when the ground state has a non-trivial sign or phase structure, the NQS must learn it explicitly. This can be difficult, but it is not a fundamental obstruction~\cite{westerhoutManybodyQuantumSign2023,schurov2025}.
At the same time, it has been shown that NQS wave functions can support entanglement structures that go beyond the standard restrictions of low-bond-dimension tensor networks~\cite{deng2017,levine2019,sharir2022neural,paul2026bound}.
In fact, the universal approximation capabilities of artificial neural networks~\cite{Cybenko1989,Hornik1991} guarantee that arbitrary wave functions can be represented as an NQS in the limit of large network sizes. This constitutes the foundation for self-consistent accuracy checks, rendering the approach systematically controlled.
While tensor networks share this property, a key advantage of NQS is that different neural network architectures can be used to confirm convergence under varying approximation bias.

In this perspective, we review recent progress in applying NQS to lattice quantum many-body problems in condensed matter.
We discuss where NQS are already competitive with alternative state-of-the-art methods, why NQS are suitable for those settings, which obstacles still limit their broader use, and what future directions appear promising at this point.
Recent reviews have introduced the general NQS framework, architectures, and applications, and we refer the reader to them for more technical discussions~\cite{medvidovicNeuralnetworkQuantumStates2024,Lange2024QST}. For current state-of-the-art NQS work in continuum electronic structure, we refer to Refs.~\cite{Pfau_2024,nys2024ab}.
The scope of the present work is to highlight the most successful NQS architectures and best practices for optimization and sampling strategies. To support these choices, we discuss recent results for frustrated spin systems, interacting fermions, and real-time dynamics that are state-of-the-art not only within NQS. Throughout, comparisons to tensor-network, quantum Monte Carlo, and dynamical mean-field approaches serve as a reference for identifying the class of problems for which NQS are most likely to extend the reach of classical simulations of strongly correlated quantum matter.

\section{Neural Quantum States in a nutshell}
\label{sec:nqs}

\begin{figure}[t]
    \centering
    \includegraphics[width=0.95\linewidth]{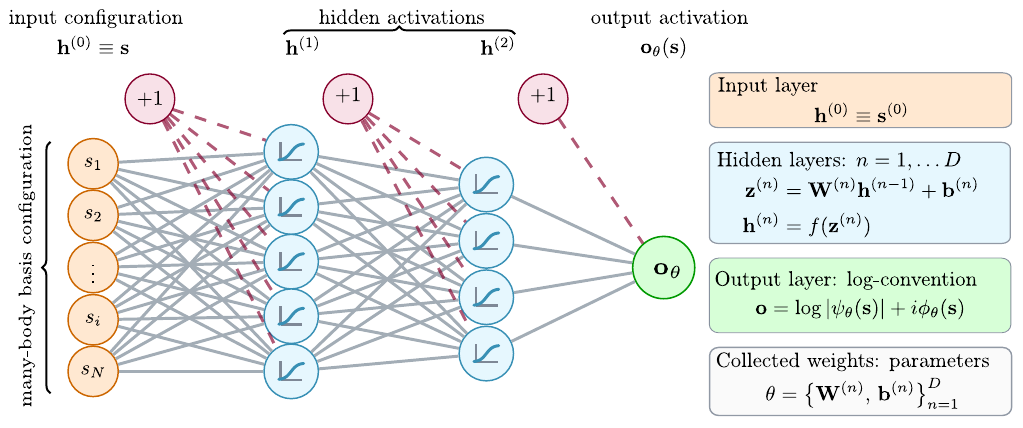}
    \caption{Schematic illustration of a feed-forward neural network, the prototypical artificial neural-network architecture used in NQS. The network maps an input configuration through a sequence of hidden layers to an output wave-function amplitude. Each layer consists of an affine linear transformation followed by an activation function $f$, which is nonlinear and typically strictly monotonic.}
    \label{fig:nqs_nutshell}
\end{figure}

In NQS, the wave-function compression is achieved by representing the wave function in a \textit{fixed local product basis}, often referred to as the \textit{computational basis}.
For instance, for a spin-1/2 system of $N$ sites one commonly uses the joint eigenbasis of the local spin-$z$ operator ($\hat S^z_i$), that is the product basis $\ket{\mathbf{s}}$ with $\mathbf{s}\in\{-1,1\}^N$. In this basis the exact wave-function amplitudes $\psi_{s_1s_2\dots s_N}=\braket{\mathbf{s}}{\psi}$ are replaced by a parameterized \textit{ansatz} $\psi_\theta(\mathbf{s})$ defining the variational state
\begin{equation}
    \label{eq:var_WF}
    \ket{\psi_\theta} = \sum_{\mathbf{s}} \psi_{\theta}(\mathbf{s})\ket{\mathbf{s}} \;.
\end{equation}
The ansatz can be understood both as a \textit{function} $\psi_\theta:\mathbf{s}\mapsto \psi_\theta(\mathbf{s})$ that maps configurations to amplitudes, and as a \textit{continuous function} $\theta\mapsto \ket{\psi_\theta}$ that maps \textit{variational parameters} $\theta\in\mathbb{R}^{N_P}$ to a quantum state, as shown in Fig.~\ref{fig:nqs_nutshell}.
Introducing the parametrized ansatz replaces the exponential storage of $\dim\mathcal H$ amplitudes by $N_P$ parameters, which is the key step toward a compressed approximation of quantum many-body states.

For NQS, this map from configurations to amplitudes is represented by an ANN.
The simplest example is the feed-forward network sketched in Fig.~\ref{fig:nqs_nutshell}, where the input configuration is transformed through successive compositions of linear and non-linear transformations into an output amplitude. This elementary construction provides the starting point for more problem-specific choices and design principles for NQS architectures, which we discuss in Sec.~\ref{sec:architecture}.

After choosing an ansatz and defining the NQS $\ket{\psi_\theta}$, one has access to any expectation value of the form
\begin{equation}
\label{eq:vmc_expectation}
\langle \hat O\rangle \equiv \frac{\bra{\psi_\theta}\hat O\ket{\psi_\theta}}{\bra{\psi_\theta}\ket{\psi_\theta}}
= \sum_{\mathbf{s}} p_\theta(\mathbf{s}) O_{\rm loc}(\mathbf{s}) ,
\end{equation}
where $p_\theta(\mathbf{s}) \equiv \tfrac{|\psi_\theta(\mathbf{s})|^2}{\sum_{\mathbf{s}} |\psi_\theta(\mathbf{s})|^2}$ is the Born distribution defined by the current variational state, and
\begin{equation}
\label{eq:local_estimator}
    O_{\rm loc}(\mathbf{s}) \equiv
    \sum_{\mathbf{s}':O_{\mathbf{s}\mathbf{s}'}\neq 0} O_{\mathbf{s}\mathbf{s}'}
    \frac{\psi_\theta(\mathbf{s}')}{\psi_\theta(\mathbf{s})}, \qquad
    O_{\mathbf s\mathbf s'} = \bra{\mathbf{s}} \hat{O} \ket{\mathbf{s}'}
\end{equation}
is the estimator or \textit{local observable} for any physical observable $\hat{O}$. The only requirement on $\hat{O}$ is that it be sufficiently sparse \cite{Nest2009SimulatingQC}, so that the sum over all $\mathbf{s}'$ with $O_{\mathbf{s}\mathbf{s}'}\neq 0$ is tractable.
The remaining sum over states $\mathbf{s}$ in Eq.~\eqref{eq:vmc_expectation} cannot in general be evaluated exactly, since the Hilbert-space dimension grows exponentially with $N$.
Instead, NQS uses \textit{Monte Carlo} (MC), replacing the sum by MC estimates obtained from samples drawn from $p_\theta$
\begin{equation}
\label{eq:mc_estimator}
\langle \hat O\rangle \approx \frac{1}{N_S}\sum_{i=1}^{N_S} O_{\rm loc}\!\big(\mathbf{s}^{(i)}\big),
\qquad \mathbf{s}^{(i)} \sim p_\theta(\mathbf{s}).
\end{equation}
The Monte Carlo estimator in Eq.~\eqref{eq:mc_estimator} is unbiased and converges to the expectation value of the \textit{variational} state $\ket{\psi_\theta}$ as the number of samples $N_S$ increases, with the statistical error decreasing as $\mathcal{O}(N_S^{-1/2})$.

Although NQS use artificial neural networks, their training is usually not data driven. In conventional supervised learning, a neural network is trained from reference input-output pairs and learns to reproduce an underlying mapping from these examples \cite{goodfellow2016}.

For quantum many-body states, this is typically not feasible, as the target wave function is itself unknown and cannot be provided as training data. Instead, one determines the network parameters through mathematical and physical constraints that the state must satisfy. The paradigmatic example is the ground state $\ket{\psi_0}$, which is singled out by the \textit{variational principle}
\begin{equation}
\label{eq:gs_energy}
E_0 = \min_{\psi} \frac{\bra{\psi}\hat H\ket{\psi}}{\bra{\psi}\ket{\psi}}\, .
\end{equation}
If the ground state is not degenerate, the minimum is unique up to an overall phase. Restricting the search to the variational family $\ket{\psi_\theta}$ gives the variational energy
\begin{equation}
    E_0 \le E(\theta) = \frac{\bra{\psi_\theta}\hat H\ket{\psi_\theta}}{\bra{\psi_\theta}\ket{\psi_\theta}}\;.
\end{equation}
Thus, optimizing $\theta$ amounts to minimizing an upper bound to the exact ground-state energy. The variational energy therefore plays the role of the \textit{loss function} \cite{goodfellow2016}, and its minimization is discussed in Sec.~\ref{sec:gs_search}.

The other widely used principle concerns the time evolution of a variational state $\ket{\psi_{
\theta_t}}$, where $\theta_t$ are the NQS parameters at time $t$, which have to satisfy the time-dependent Schr\"odinger equation up to the residual $\ket{\phi}$
\begin{equation}
\label{eq:time_evolution_residual}
\ket{\phi} = i \sum_j [\dot{\theta}_t]_j \ket{\partial_j \psi_{\theta_t}} - \hat{H} \ket{\psi_{
\theta_t}}\,, \qquad \text{with}\; \bra{\partial_j \psi_{
\theta_t}}\ket{\phi} - \frac{\bra{\partial_j \psi_{
\theta_t}}\ket{\psi_{\theta_t}}\bra{\psi_{\theta_t}}\ket{\phi}}{\bra{\psi_{\theta_t}}\ket{\psi_{\theta_t}}} = 0 \;.
\end{equation}
Thus, the Schr\"odinger equation cannot usually be satisfied exactly, because $\hat{H} \ket{\psi}$ does not necessarily lie completely in the tangent space $\mathrm{span}\big\lbrace \ket{\partial_j\psi_{\theta_t}}\big\rbrace$. With the condition that the residual is completely outside of the tangent space one arrives at the \textit{time-dependent variational principle} (TDVP)
\begin{equation}
    \label{eq:tdvp_equation}
    \sum_j [\dot{\theta}_t]_j \bra{\partial_i \psi_{\theta_t}}\hat{Q}\ket{\partial_j \psi_{\theta_t}} = -i\bra{\partial_i \psi_{\theta_t}} \hat{Q}\hat{H} \ket{\psi_{\theta_t}}\, , \qquad \text{with}\,\hat{Q} = 1 - \frac{\ket{\psi_{\theta_t}}\bra{\psi_{\theta_t}}}{\bra{\psi_{\theta_t}}\ket{\psi_{\theta_t}}} \;,
\end{equation}
which allows the propagation of NQS in time. The matrix $S_{ij}\equiv\bra{\partial_i \psi_{\theta_t}}\hat{Q}\ket{\partial_j \psi_{\theta_t}}$ on the left-hand side of Eq.~\eqref{eq:tdvp_equation} is the \textit{quantum geometric tensor} (QGT) and the vector $F_i\equiv-i\bra{\partial_i \psi_{\theta_t}} \hat{Q}\hat{H} \ket{\psi_{\theta_t}}$ on the right-hand side is the \textit{force vector}.

The key property of Eqs.~\eqref{eq:gs_energy} and~\eqref{eq:tdvp_equation} is that, for local Hamiltonians, the energy, its gradients, the force vector and QGT can be written as MC expectation values over the Born distribution $|\psi(\mathbf{s})|^2$. This makes it possible to optimize $E(\theta)$ with gradient-based methods and to evaluate the parameter time derivatives required by TDVP (as discussed in more detail in Sec.~\ref{sec:gs_search} and Sec.~\ref{sec:dynamics}). In this respect, NQS optimization resembles the minimization of loss functions in machine learning, although it operates without reference data. In practice, this Variational Monte Carlo (\textit{VMC})---or \emph{tVMC} in the context of time evolution---formulation is what makes both ground-state search and time evolution feasible in Hilbert spaces too large for explicit summation.

\section{The state of the art}
In this section we discuss examples of NQS applied to condensed matter problems, obtaining state-of-the-art numerical results when compared to other methods.
This illustrates the types of problems where NQS can overcome the limitations inherent in other numerical approaches.

\subsection{Ground states of frustrated magnets}
\label{sec:j1j2}
\begin{figure}
\centering
\includegraphics[width=10cm]{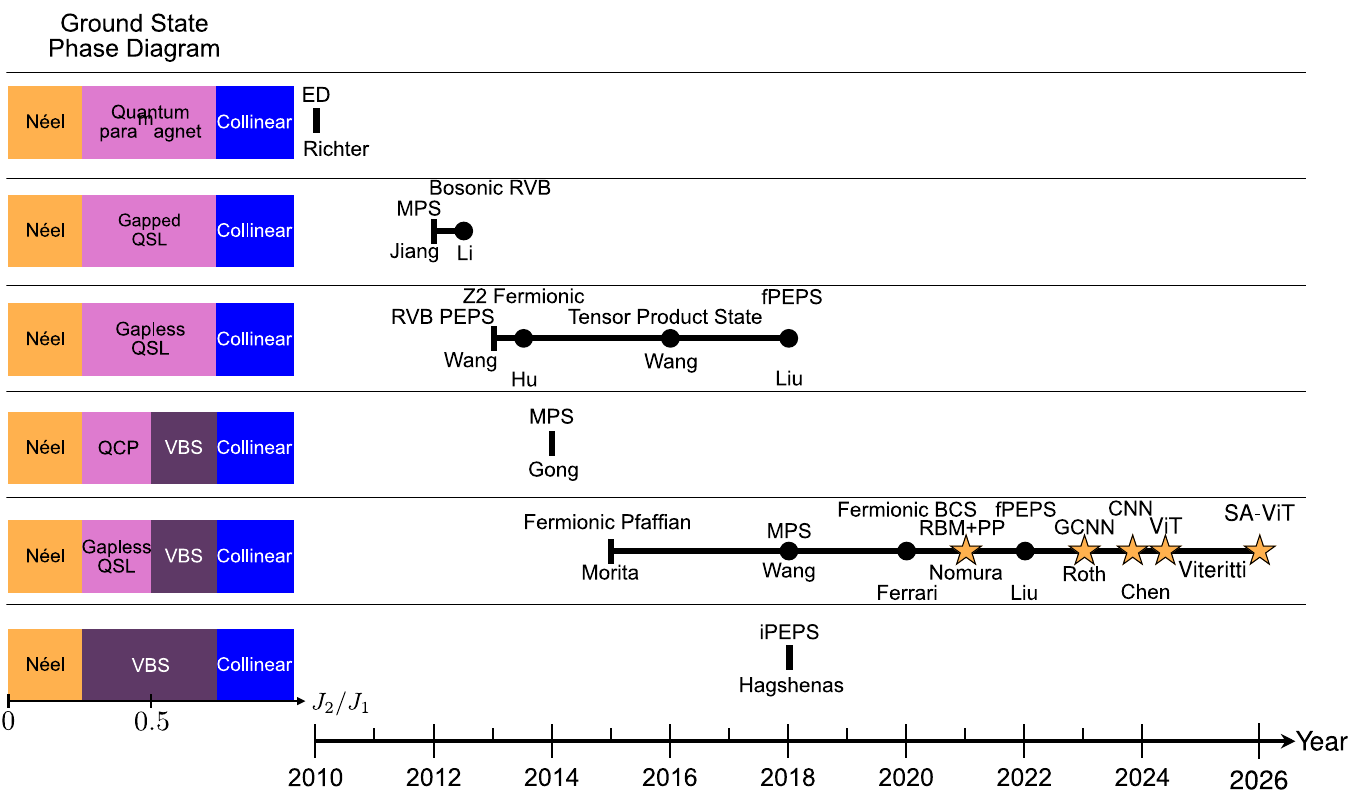}
\includegraphics[width=4cm]{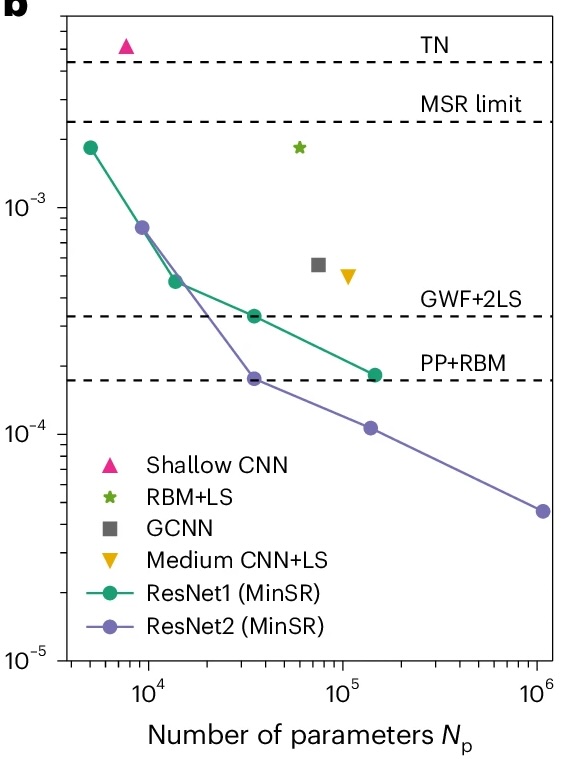}
\caption{
Numerical results for the ground state of the $J_1-J_2$ model.
(Left) Timeline of ground state phase diagrams using various numerical methods since $2010$.
The type of method / ansatz is indicated along with the lead author of each study.
NQS results are indicated by stars.
The works shown are compiled from Refs.~\cite{richter2010,jiang2012,li2012bosonicrvb,wang2013rvbpeps,hu2013,gong2014,morita2015,wang2016tps,haghshenas2018,liu2018,wang2018,ferrari2020,nomura2021,roth2023,chen2024,rende_2024,viteritti2026thermodynamiclimitnqs}.
A consensus has been reached that the ground state phase diagram consists of Néel, gapless spin liquid, valence bond solid and collinear phases.
(Right) Comparison of variational energies for $J_2/J_1 = 0.5$ using different NQS ans\"atze (adapted from Ref.~\cite{chen2024} under~\CCBYFour).
}
\label{fig:j1j2}
\end{figure}
Our first example is the study of ground states of frustrated quantum magnets~\cite{lhuillier2011}, where NQS have recently developed into a powerful tool.
Frustrated quantum magnets have been well studied over the past few decades as systems which may host exotic states of matter such as quantum spin liquids~\cite{zhou2017,savary2017,knolle2019,broholm2020}.
The success of NQS is best illustrated by work on the $J_1-J_2$ model on the square lattice,
\begin{equation}
    \hat{H} = J_1 \sum_{\langle ij \rangle} \hat{\mathbf{S}}_i \cdot \hat{\mathbf{S}}_j + J_2 \sum_{\langle \langle ij \rangle \rangle} \hat{\mathbf{S}}_i \cdot \hat{\mathbf{S}}_j\,,
\end{equation}
where $J_1$ couples nearest and $J_2$ next-nearest neighbors.
In this and many other condensed-matter Hamiltonians, the matrix elements of the Hamiltonian are real, and the off-diagonal elements have inhomogeneous signs in the sampling basis $\lbrace \ket{\mathbf{s}}\rbrace$, so the ground state has real components with a non-trivial sign structure.
Therefore, the components $\psi(\mathbf{s}) = \vert\psi(\mathbf{s})\vert e^{i\phi(\mathbf{s})}$ of the exact ground state will have quantum phases $\phi(\mathbf{s}) \in \{0,\pi\}$, and the ansatz should not only approximate the magnitude $\abs{\psi(\mathbf s)}$, but also the non-trivial sign structure, $e^{i\phi(\mathbf{s})} \in \{-1,+1\}$.

For antiferromagnetic, non-zero $J_1, J_2$, quantum Monte Carlo has a sign problem and the Marshall sign rule~\cite{marshall1955} does not solve the ground-state sign structure.
Furthermore, around $J_2/J_1 = 0.5$, competing ground-state orders make variational approaches challenging.

Despite efforts using various methods, including density matrix renormalization group (\textit{DMRG})~\cite{jiang2012}, conventional \textit{VMC}~\cite{hu2013} and projected entangled pair states (\textit{PEPS})~\cite{haghshenas2018, liu2018}, the ground-state phase diagram in the quantum paramagnetic regime (around $J_2/J_1 = 0.5$), remained controversial (see Fig.~\ref{fig:j1j2}).
One possible scenario~\cite{gong2014, morita2015, wang2018, ferrari2020} was that the region consisted of two phases: a gapless quantum spin liquid and a valence bond solid.
However, resolving whether the spin liquid was indeed a phase or a quantum critical point required precise finite-size scaling, with indirect evidence coming from the locations of level crossings~\cite{wang2018,ferrari2020}.
Nomura and Imada~\cite{nomura2021} used a \textit{RBM+PP} wave function to simulate up to $18 \times 18$ lattices with periodic boundaries, obtaining the most accurate ground-state solutions yet.
The RBM+PP wave function constructs an ansatz
\begin{equation}
    \psi_{\theta}^{\mathrm{RBM+PP}}(\mathbf s) = \mathcal{J}_{\theta}^{\mathrm {RBM}}(\mathbf{s})\psi_{\theta}^{\mathrm {PP}}(\mathbf{s})
\end{equation}
from the combination of a shallow neural network (a restricted Boltzmann machine~\cite{carleoSolvingQuantumManybody2017b}), $ \mathcal{J}_{\theta}^{\mathrm {RBM}}(\mathbf{s})$, and a more conventional pair-product wave function, $\psi_{\theta}^{\mathrm {PP}}(\mathbf{s})$~\cite{nomura2017}.
The authors also enforced symmetries on the ansatz in order to obtain accurate results.
They showed that the level crossings did indeed correspond to the claimed phase transitions and with careful finite-size scaling of the order parameters and correlation ratios confirmed this ground-state phase diagram.
Nevertheless, the \textit{RBM+PP} wave function relies on a parton mean-field decoupling of the Hamiltonian to derive the pair-product part, therefore introducing a source of physical bias.
Similar ground-state energies were later obtained using a group-convolution neural network (\textit{GCNN})~\cite{roth2023}, retaining the role of symmetries while removing the mean-field-derived component of the wave function.
Subsequently, the introduction of \textit{minSR}~\cite{chen2024} (see Sec.~\ref{sec:gs_search}) allowed the use of much larger neural networks of up to $\mathcal{O}(10^6)$ parameters, resulting in even more accurate ground states with translationally-equivariant neural networks~\cite{chen2024, rende_2024, viteritti2026thermodynamiclimitnqs}.
This constituted a breakthrough in terms of tackling ground states with non-trivial sign structure in two dimensions, using wave functions free from mean-field bias and applicable to systems with periodic boundary conditions.

As a result, these neural network architectures have been successfully applied to study the ground states of other models, such as the Shastry-Sutherland model~\cite{viteritti2025}.

\subsection{Ground States of the Hubbard Model}
\begin{figure}
    \centering
    \includegraphics[width=\linewidth]{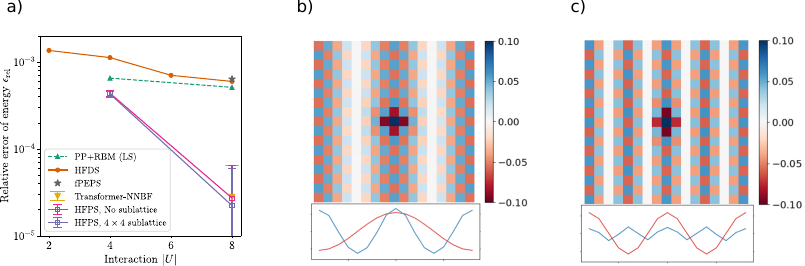}
    \caption{a) Energies reached on $8\times 8$ for various architectures at half filling. b-c) Stripe pattern at $\frac{1}{8}$ hole doping for only nearest neighbor hopping (b) and including next to nearest neighbor hopping. The top panel shows the spin correlation function, while the bottom panel shows cuts of the staggered spin correlation function (red) and density (blue). All panels are adapted from Ref.~\cite{chen2025} and published under~\CCBYFour.}
    \label{fig:hubbard}
\end{figure}
In the study of itinerant, interacting lattice fermions, the square-lattice Hubbard model
\begin{equation}
    \hat H = - t\sum_{\expval{ij}, \sigma} \big( \hat c^\dag_{i\sigma} \hat c_{j\sigma} + \hat c^\dag_{j\sigma} \hat c_{i\sigma}\big) + U \sum_i \hat n_{i\up} \hat n_{i\down},
\end{equation}
gives rise to a rich variety of physical phenomena from simple microscopic rules, including qualitative features of cuprate superconductors \cite{keimerQuantumMatterHightemperature2015c}.
It consists of a kinetic term, with hopping $t > 0$, and an onsite interaction term of strength $U$.
A particularly numerically challenging problem is finding the ground state of the hole doped square lattice model with filling fraction around $n=0.875$ and repulsive, moderately large interaction $U$ as \textit{QMC} solutions are hindered by the sign problem~\cite{maier2005,scalapino2007} and various near degenerate low-energy solutions make the problem also hard to deal with variationally~\cite{leblanc2015, PhysRevX.10.031016, Darmawan2018, Zheng2017}.
By now a consensus on length $\lambda = 8$ filled stripes as the ground state has been reached in this regime by various numerical techniques~\cite{Zheng2017, ido2018, tocchio2019, comp_persepective_HU_qin}.

More recently, NQS were applied to this part of the ground-state phase diagram~\cite{gu2025, chen2025, roth2025, rende_superconductivity_2026}, achieving the lowest variational energies yet, with clear signatures of the $\lambda = 8$ striped phase, on large lattices of up to $16 \times 16$ with periodic boundary conditions.

Previous work on the $t-t'$ Hubbard model with next nearest neighbor hopping $t'$ suggests that the ground state at $t'/t = -0.2$, $U/t = 8$ and $n = 0.875$ is a partially filled superconducting state~\cite{xu2024} in coexistence with spin density stripes with length $\lambda$ fluctuating around $\lambda = 4$ depending on the system size.
Gu et al.~\cite{gu2025} focused on the details of the stripes using NQS, showing that this fluctuation, along with the crossover in stripe direction, were boundary effects which could be eliminated using periodic boundary conditions.
The ability of NQS to simulate genuinely two-dimensional systems with periodic boundaries offers a major advantage over other methods, allowing one to identify boundary effects at smaller system sizes.

Using a different NQS architecture based on Pfaffians, a superconducting $\lambda = 4$ partially-filled stripe phase was found~\cite{roth2025}, qualitatively consistent with the results in Ref.~\cite{xu2024}. Despite various works showing comparable variational energies for the groundstate in this regime~\cite{gu2025, chen2025, roth2025, rende_superconductivity_2026}, they do not agree on the nature of the ground state. Viteritti~\textit{et al.} reported convergence between different NQS architectures after both symmetry restoration and energy-variance reduction~\cite{viteritti2026variationalbiasresolvingintertwined}. Rende~\textit{et al.} introduced a symmetry-preserving ansatz intended to avoid broken-symmetry local minima~\cite{rende_superconductivity_2026}. They achieved the lowest variational energies up to date for systems up to $24\times 24$, finding superconducting order.

\subsection{Dynamics}
\begin{figure}
    \centering
    \includegraphics[width=\linewidth]{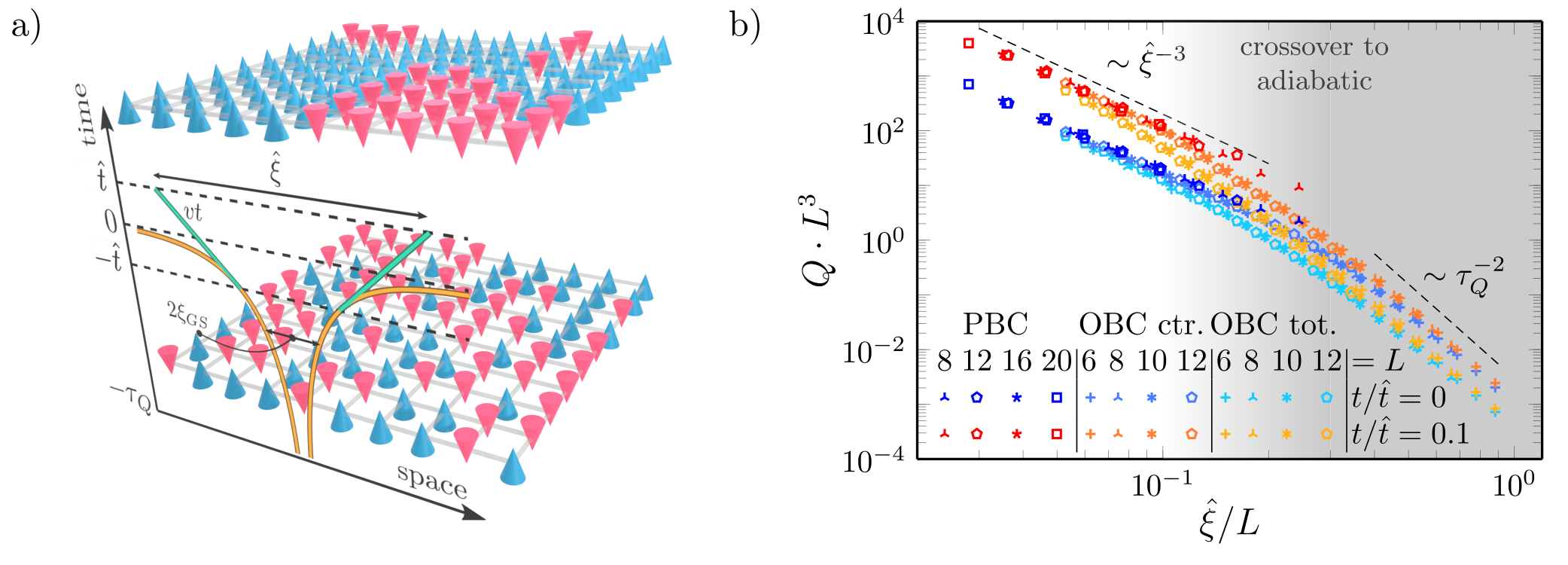}
    \caption{The schematic a) shows a space-time picture of a quench across a critical point at $t = 0$. During a ramp parameterized by the dimensionless distance to the critical Hamiltonian parameter $\varepsilon(t) = t/\tau_Q$, critical slowing down prevents the correlation length from following the equilibrium divergence of the correlation length, so $\xi$ saturates at a finite freeze-out scale $\hat{\xi}$. A freeze-out length scale together with the freeze-out time $\hat{t}$ form a \textit{sonic horizon}, so that the system breaks symmetry in quasi-independent domains of size $\hat{\xi}$, producing \textit{defects} and thus \textit{excitations}. Figure b) summarizes the results of the finite size examination of the quantum Kibble-Zurek mechanism in Ref.~\cite{schmittQuantumPhaseTransition2022}. It shows the finite-size scaling of the excitation energy density $Q = \frac{1}{L^2}\big(\langle\hat{H}(t)\rangle - E_0(t)\big)$ at the critical point $t = 0$, shown as $Q L^3$ vs $\hat{\xi}/L$ for different $L (=\sqrt{N})$ and boundary conditions. For $\hat{\xi} \ll L$, the data collapse follows the \textit{quantum Kibble-Zurek} prediction $Q\propto \hat{\xi}^{-(d + z)} = \hat{\xi}^{-3}(d = 2, z = 1)$. As $\hat{\xi}$ approaches $L$, the finite-size gap induces a crossover to an adiabatic regime with $Q \propto \tau^-2_Q$. The collapse across periodic (NQS) and open-boundary tensor-network simulation demonstrates universality beyond ramp protocol and boundary details, and the system sizes reached in the NQS simulations are essential to identify the scaling behavior in the non-adiabatic regime. Both figures are reproduced from Ref.~\cite{schmittQuantumPhaseTransition2022} and published under~\CCBYFour.}
    \label{fig:kzm}
\end{figure}

While the ground state encodes the static zero-temperature properties of a Hamiltonian, out-of-equilibrium dynamics probes a much larger portion of its spectrum. Access to dynamics therefore reveals quasiparticles and collective modes through response functions \cite{damascelliAngleresolvedPhotoemissionStudies2003,landigMeasuringDynamicStructure2015}, tests the stability of ergodicity-breaking phases \cite{abaninManybodyLocalizationThermalization2019,deroeckStabilityInstabilityDelocalization2017}, and can even expose intrinsically dynamical phases of matter such as time crystals \cite{wilczekQuantumTimeCrystals2012,zhangObservationDiscreteTime2017}.

This importance is matched by the numerical difficulty, as real-time evolution is among the hardest problems in many-body physics.
The introduction of NQS opened a promising new route to this long-standing problem. Early time-evolution studies based on the TDVP formulation of Eq.~\eqref{eq:tdvp_equation} already demonstrated the viability of this approach for the transverse-field Ising model (TFIM) in one dimension. The substantially more demanding two-dimensional TFIM,
\begin{equation}
    \hat H = -J \sum_{\langle ij\rangle} \hat S_i^z \hat S_j^z - h \sum_i \hat S_i^x \, ,
\end{equation}
is of greater physical interest and has therefore become a particularly useful benchmark. In an early benchmark, Schmitt and Heyl showed that NQS can simulate quench dynamics in this model on lattices as large as $10\times10$, reaching timescales comparable to or beyond \textit{then}-state-of-the-art tensor-network methods
\cite{schmitt2020quantum}. The techniques introduced in Ref.~\cite{schmitt2020quantum} were subsequently used, with problem-specific modifications, to compute the spectral function of the TFIM close to criticality \cite{mendes-santos2023}. This unprecedented access to late-time dynamics also enabled a clear numerical test of the quantum Kibble--Zurek mechanism through a combination of TN and NQS simulations, again in the TFIM \cite{schmittQuantumPhaseTransition2022}. In this picture, a finite-rate sweep across a quantum critical point becomes nonadiabatic because of critical slowing down, generating defects in the symmetry-broken phase, here ferromagnetic domains separated by domain walls. Ref.~\cite{schmittQuantumPhaseTransition2022} thereby confirmed the predicted finite-size scaling of the excitation-energy density with the correlation length, as illustrated in Fig.~\ref{fig:kzm}.
Recent work extended this approach to the case of the three-dimensional quantum Ising model with up to 1000 qubits \cite{Naik20262}.

In a similar vein, Medvidovi\'c et al. computed the Loschmidt echo in the two-dimensional quantum rotor model on systems of up to $8\times8$ sites, showing that NQS combined with Hamiltonian Monte Carlo yields stable unitary real-time dynamics and quantitatively accurate echoes \cite{medvidovic2023variational}.

Beyond standard TDVP integration, an important alternative is to propagate the state over a short time step and then variationally project the result back onto the ansatz manifold $\mathcal{M}=\{\ket{\psi_\theta}\,|\,\theta\in\mathbb{R}^{N_P}\}$. In \textit{projected time-dependent variational Monte Carlo} (p-tVMC), this projection is formulated as an optimization problem at each time step, for example by minimizing the infidelity between $\ket{\psi_{\theta_{t+1}}}$ and the propagated state $e^{-i\delta t H}\ket{\psi_{\theta_t}}$ \cite{sinibaldiUnbiasingTimedependentVariational2023}. This is particularly useful when the wave function develops exact or approximate zeros, where standard TDVP-based propagation can struggle \cite{sinibaldiUnbiasingTimedependentVariational2023}. Closely related projection-based ideas have also been successfully applied to \textit{ab-initio} electron dynamics by Nys et al. \cite{nys2024ab}.

A second alternative abandons stepwise propagation altogether and instead optimizes the full trajectory at once. In the explicitly time-dependent t-NQS approach, the network represents $\psi_\theta(\mathbf{s},t)$ directly and its parameters are trained over a whole time interval by minimizing a global residual (Eq.~\eqref{eq:time_evolution_residual}) of the Schr\"odinger equation \cite{walleManybodyDynamicsExplicitly2024}. Closely related in spirit, the \textit{time-dependent neural Galerkin} method of Ref.~\cite{sinibaldiTimeDependentNeuralGalerkin2026} represents the trajectory as a time-dependent linear combination of time-independent NQS basis states and optimizes it through a global-in-time variational principle. Both approaches have been benchmarked successfully on the TFIM, including long-time dynamics.

Open quantum systems constitute another established application for dynamic NQS methods. The unitary evolution is replaced by a Lindblad master equation for the density operator. Early work introduced neural density operators based on latent-space purification, restricted Boltzmann machine density matrices and purified steady-state ans\"atze, and used them to compute non-equilibrium steady states or dissipative dynamics of spin models \cite{torlai2018latent,hartmann2019neural,nagy2019variational,vicentini2019,Yoshioka_2019}. The derivation of a TDVP directly for Lindblad evolution with autoregressive neural density operators enabled real-time dissipative simulations of one- and two-dimensional Heisenberg systems as well as confinement dynamics with losses \cite{reh2021time}. More recent formulations either represent the open-system state through measurement probabilities in a POVM basis or employ deeper density-operator architectures, including Liouville-space, convolutional, and transformer ans\"atze, broadening the class of accessible Lindbladian steady states \cite{Luo2022,kothe2024liouville,mellak2024deep,wei2025variational}. Related applications now extend to lattice-gauge and quantum-optical settings, including the Schwinger model, waveguide-QED emitter arrays, and many-body super- and subradiant light--matter dynamics \cite{Lin_2024,Vovk2026,lagnese2026neural}.

\section{Best Practices}

Over recent years, several best practices for applying NQS have been identified, mostly empirically:
\begin{enumerate}
    \item \textbf{Symmetries.} Enforce as many symmetries as possible (Sec.~\ref{sec:bias}).
    \item \textbf{Spin architectures.} For spin models, use a translationally-equivariant or group-equivariant architecture such as a CNN or GCNN (Sec.~\ref{sec:architecture}).
    \item \textbf{Fermionic signs.} For fermionic models, use a determinant- or Pfaffian-based ansatz (Sec.~\ref{sec:architecture}).
    \item \textbf{Ground states.} For ground-state calculations, optimize the wave function using minSR, enabling the use of deep neural networks (Sec.~\ref{sec:gs_search}).
    \item \textbf{Dynamics.} For dynamics with TDVP, use as many samples as possible to estimate the QGT and regularize its spectrum carefully for the inversion (Sec.~\ref{sec:dynamics}).
    \item \textbf{Parallelization.} Use GPUs to parallelize sampling and observable estimation (Sec.~\ref{sec:gpu}).
\end{enumerate}
In the following, we discuss these points in more detail.

\subsection{Physical Bias}
\label{sec:bias}
The central idea of NQS is to use asymptotically unbiased and expressive wave functions such that an accurate approximation of the true solution can be found via optimization, in a wide range of different phases of matter.
However, using very unstructured wave functions makes finding a nearly-optimal solution difficult~\cite{reh2023}, so some physical bias should be included to guide the optimization towards physically-relevant solutions.
Some of these are exact constraints, such as symmetry, or sign structure in the cases when they are exactly known.
Others are more general notions of the functional dependence of the wave function on its input variables, such as locality.
In this subsection, we discuss aspects of these physical biases which are important factors in NQS design.

First, since condensed matter Hamiltonians in most cases exhibit local interactions, this leads to the notion of locality in their wave functions, for example in spin systems~\cite{Chen_2023, Lieb1972, hastings2010localityquantumsystems} and, from an entanglement perspective, the area law for non-critical systems~\cite{hastings2010localityquantumsystems, Eisert2010}.
The need to work with local structures arises not only in \textit{NQS}, but also naturally in a wide range of machine-learning tasks, including image classification, time-series analysis, and natural language processing. Accordingly, a variety of architectures originally developed for such applications, including \textit{CNNs}~\cite{goodfellow2016cnn}, \textit{RNNs}~\cite{goodfellow2016}, and \textit{GCNNs}~\cite{cohen2016}, have also found use as NQS. We discuss these architectures in more detail in Sec.~\ref{sec:architecture}.

Next, condensed matter Hamiltonians possess symmetries, which means that eigenstates must belong to irreducible representations (irreps) of the symmetry group~\cite{el-batanouny2008}.
Enforcing this structure explicitly can improve the accuracy of variational energies and other observables~\cite{tahara2008, viteritti2026variationalbiasresolvingintertwined,morita2015,nomura2021a}.
A recent example is provided by Viteritti~\textit{et al.}~\cite{viteritti2026variationalbiasresolvingintertwined}, who compared Pfaffian- and determinant-based states with nearly identical variational energies but different correlation functions, especially in the pairing channel. Including translational and rotational symmetries substantially reduced these discrepancies, illustrating that reliable physical observables may require symmetry enforcement even when the variational energies already agree.

In particular, for discrete symmetries, this can be done via a quantum number projection of the form
\begin{equation}
    \psi_{\mathrm{symm}}(\mathbf{s}) = \frac{d_\alpha}{\abs{G}} \sum_{g \in G} \chi_g^* \psi (g^{-1} \mathbf{s}),
    \label{eq:symm}
\end{equation}
where $G$ is the symmetry group to be symmetrized over, $d_\alpha$ the irrep dimension and characters $\chi_g$ corresponding to the group element $g$, projecting the wave function to a desired irreducible representation.
Typically, Eq.~\eqref{eq:symm} is used to enforce lattice symmetries (the space group), as well as discrete spin symmetries, such as spin-parity symmetry in the computational basis.
However, as the number of translation group elements scales with the number of lattice sites, naively applying this quantum number projection is prohibitively expensive for large lattices.
This limitation can be overcome by a judicious choice of neural network architecture, which we discuss in Sec.~\ref{sec:architecture}.

Beyond projection, symmetries such as particle number or magnetization can be easily enforced via sampling~\cite{nomura2021, roth2023, chen2024, viteritti2025}.
Other approaches such as enforcing SU(2) symmetry by working in a total momentum basis~\cite{Vieijra2020,luo2023gauge}, or translational symmetries via use of representative samples~\cite{choo2018, bukov2021} or through restricting sampling to a particular set~\cite{zhang2025}, have been proposed, although they have not been widely adopted.

Learning the sign structure (see Sec.~\ref{sec:j1j2}) can be particularly challenging for NQS~\cite{westerhout2020, szabo2020neural}.
For this reason, it has been common to bias trial states towards known proxy sign structures for spin systems~\cite{Choo2022, nomura2021, nomura2017}. For lattice fermions, successful wave functions are based on determinants and Pfaffians~\cite{lou2019,stokes2020,moreno2022,gauvinndiaye2025,lange2025, gu2025, chen2025}, biasing towards mean field sign structures.
From this point of view, approaches that minimize built-in bias while retaining high accuracy are preferable.

\subsection{Architectures}
\label{sec:architecture}
\begin{figure}
\centering
\includegraphics[width=0.8\textwidth]{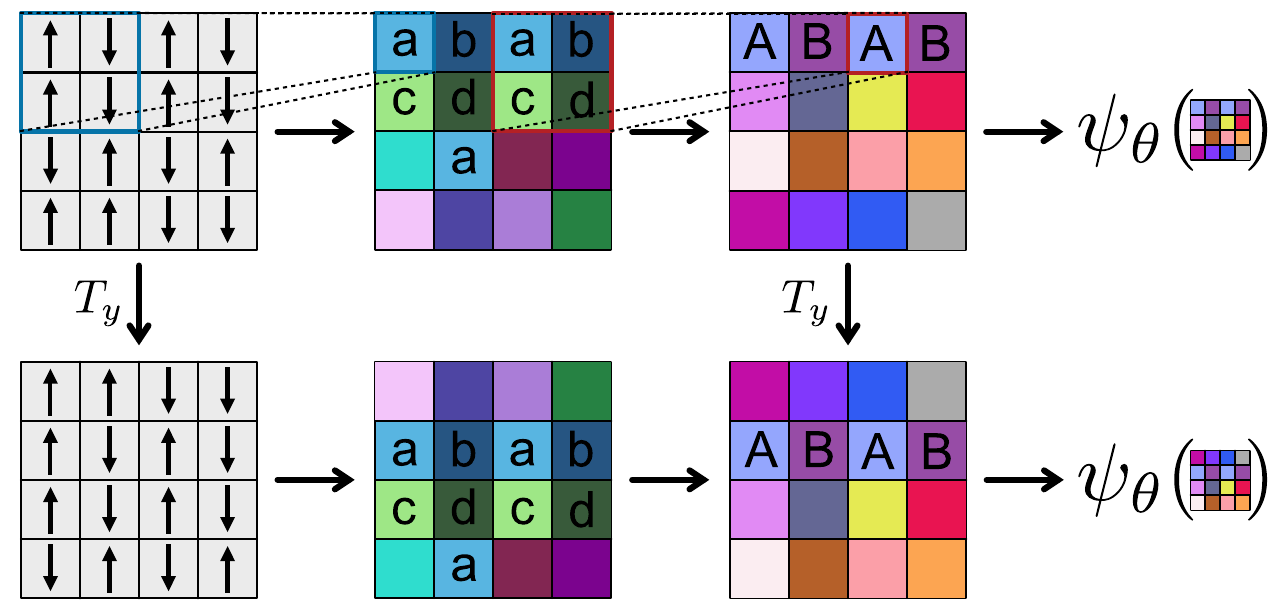}
\caption{A local, translationally-equivariant architecture.
Since the output of each layer depends on local information, locally identical environments are mapped to the same features (labelled cells).
Since each layer is translationally-equivariant, translating the output is equivalent to translating the input (compare top and bottom rows).
Therefore the wave function can be projected to a desired translational irreducible representation in a single evaluation by appropriate choice of output function.
}
\label{fig:architecture}
\end{figure}

A good choice of NQS architecture for a specific problem incorporates the physical biases discussed in the previous section in an efficient way.
Fortunately, some of these requirements are encountered in well-established machine learning applications, so neural network components developed in these contexts can be directly applied in NQS.

Both locality and translational symmetry arise in computer vision tasks, resulting in the development of deep convolutional neural networks (CNNs)~\cite{goodfellow2016cnn, cnns, krizhevsky2012}.
These are built out of convolutional layers which, for a two-dimensional system, perform the operation
\begin{equation}
\begin{aligned}
    z_{ijm}^{(n)}
    &=
    b_m^{(n)}
    + \sum_{\delta_x \in \mathcal K_x}
      \sum_{\delta_y \in \mathcal K_y}
      \sum_{c=1}^{C_{n-1}}
      K_{\delta_x\delta_y c m}^{(n)}
      h_{i+\delta_x,j+\delta_y,c}^{(n-1)}, \\
    h_{ijm}^{(n)}
    &= f\!\left(z_{ijm}^{(n)}\right).
\end{aligned}
\end{equation}
Here $h^{(0)}\equiv\mathbf{s}$ is the input configuration, $m=1,\dots,C_n$ labels output channels, and $\mathcal K_x,\mathcal K_y$ are the spatial kernel offsets of extents $k_x$ and $k_y$.
The convolutional kernel $K^{(n)}\in \mathbb{R}^{k_x \times k_y \times C_{n-1} \times C_n}$ is the local, weight-shared analogue of the dense matrix $W^{(n)}$ of the linear transformation in Fig.~\ref{fig:nqs_nutshell}, with bias $b_m^{(n)}$; spatial indices are interpreted using the boundary convention of the lattice, for example modulo $N_x,N_y$ for periodic boundaries.
Finally, the nonlinearity $f$ acts element-wise on its input.

These layers incorporate locality as the output at a given position $(i,j)$ depends only on the input variables in a given region around it (see Fig.~\ref{fig:architecture}).
In order to exploit this, the input to convolutional layers in an NQS~\cite{liang2018,choo2019,schmitt2020quantum,reh2023} should be shaped with the locality of the problem in mind, for example ensuring that the positions of variables in $x$ are related to their positions on the lattice, and using \emph{patching} to ensure up and down spin electrons~\cite{chen2025,sharma2025} or variables within a unit cell~\cite{viteritti2025} are associated with the same spatial position.

Convolutional layers are also translationally equivariant~\cite{cohen2016, bronstein2021},
\begin{equation}
    y(T_g x) = T_{g'} y(x),
\end{equation}
where on the left-hand side a translation, $T_g$, is applied to the input and on the right-hand side, $T_{g'}$, to the output of the convolutional layer (see Fig.~\ref{fig:architecture}).
Using Eq.~\eqref{eq:symm}, an NQS constructed from translationally-equivariant layers can be projected to a momentum sector (translation group irrep) during a single network evaluation.
The wave function can be further symmetrized over the point group by quantum number projection, projecting the wavefunction efficiently to a specific irrep of the lattice space group.
CNNs can be generalized to be symmetric under the full space-group in GCNNs~\cite{cohen2016,roth2021}, projecting to a space group irrep in a single evaluation.

Combined with the development of minSR (see Sec.~\ref{sec:optimization}), allowing optimization of neural networks with large numbers of parameters, state-of-the-art deep translationally-equivariant NQS for ground states of spin problems have been developed~\cite{viteritti2023,chen2024,viteritti2025,viteritti2026thermodynamiclimitnqs,nutakki2025}, inspired by highly optimized architectures used for computer vision tasks~\cite{he2015,dosovitskiy2021,liu2022convnet}.
These are either explicitly convolutional networks~\cite{chen2024} or vision transformers with factored or spatial attention~\cite{viteritti2023,viteritti2026thermodynamiclimitnqs}, which have broadly similar structures~\cite{nutakki2025}.
The structure of these networks consists of an embedding layer, used to set the translational unit cell of the NQS, a deep encoder which extracts a high-dimensional hidden representation~\cite{viteritti2025}, and a shallow, complex output block.

For fermions, the advantage of using lattice symmetry-equivariant networks is not as clear-cut as for spins.
As discussed in Section~\ref{sec:bias}, high-accuracy fermionic NQS use the output of neural networks as the elements of a determinant or Pfaffian state.
While it is possible to use the translationally-equivariant architectures described above with determinants~\cite{Romero2025} and Pfaffians~\cite{chen2025, rende_superconductivity_2026}, the equivariant encoder appears to be performing worse than symmetrization via quantum number projection (Eq.~\ref{eq:symm})~\cite{sharma2025}. Nonetheless, this approach becomes unfeasible for large systems, due to the scaling of the number of the number of elements in the space group symmetry group, tipping the scale again towards equivariant architectures~\cite{rende_superconductivity_2026}. Instead of enforcing the full lattice symmetries via quantum number projection, Chen~\textit{et al.} only restored sublattice symmetries instead in order to overcome the computational cost of quantum number projection achieving state of the art energies~\cite{chen2025}.
On the other hand, in Ref.~\cite{gu2025} the authors were able to obtain state-of-the-art energies using a neural backflow transformer without enforcing any lattice symmetries, highlighting the fact that there is more work to be done in understanding the most effective way to implement lattice symmetries for these systems.

For dynamics, state-of-the-art results have been achieved using CNNs~\cite{schmitt2020quantum,medvidovic2023variational}; however, as discussed in Section~\ref{sec:optimization}, the time-evolution algorithm restricts the size of neural networks which can be used.
Therefore, lightweight versions of the translationally-equivariant NQS used for ground states can be applied to dynamics~\cite{gravinaNeuralProjectedQuantum2025a,chen2025cwtf}.

The notion of locality can alternatively be included using autoregressive architectures~\cite{goodfellow2016}, which model the output as
\begin{equation}
\label{eq:autoregressive_network}
    \psi_{\theta}(\mathbf{s}) = \sqrt{\abs{\psi_{\theta}(s_1)}^2 \abs{\psi_{\theta}(s_2|s_1)}^2\dots \abs{\psi_{\theta}(s_N|s_1, \dots, s_{N-1})}^2}e^{i\phi(\mathbf{s})},
\end{equation}
where the conditional distributions $\abs{\psi_{\theta}(s_i|s_1,\dots,s_{i-1})}^2$ are computed in a spatially sequential fashion starting from $\psi_{\theta}(s_1)$.
Examples of autoregressive NQS include recurrent neural networks (RNNs)~\cite{hibat-allah2020} and transformers with masked self-attention~\cite{zhang2023, sprague2024}.
They have the advantage that uncorrelated samples can be drawn efficiently and directly from the wave function, rather than being generated through an MCMC process, as discussed in Sec.~\ref{sec:optimization}.
For spin models, the use of autoregressive architectures has largely been restricted to models where the ground state is real and positive (e.g., ground states of stoquastic Hamiltonians)~\cite{hibat-allah2020, hibat-allah2023, zhang2023, sprague2024, moss2025a}.
This difficulty in learning non-trivial sign structures could stem from the lack of expressivity of autoregressive wave functions~\cite{bortone2024}.

Other types of networks have been developed to incorporate problem-specific biases, such as gauge symmetries for lattice gauge theories~\cite{luo2021,luo2022gauge,favoni2022,luo2023gauge} and quantum spin liquids~\cite{kufel2025approximately}.

\subsection{Optimization and Sampling}
\label{sec:optimization}
We now turn to the algorithmic question of how to \emph{optimize} these states for ground-state searches and how to \emph{propagate} them in time.
While the underlying variational principles are conceptually the same as in traditional VMC, the practical setting is remarkably different. Modern NQS ans\"atze contain far more parameters and exhibit greater expressive power than the
strong physically biased
wave functions historically used in VMC, which often involve only $\mathcal{O}(N)$ variational parameters.
Their greater expressive power broadens the class of states that can be represented, but it also makes optimization more demanding, increases memory requirements, and often requires specialized hardware for efficient training and evaluation.
These challenges have motivated the development of algorithms tailored to large neural-network ansätze, often drawing directly on the software ecosystem and optimization strategies of modern machine learning. In this section, we discuss the resulting best practices for ground-state optimization and time evolution, as well as the sampling procedures used to estimate the quantities required for both.

\begin{figure}[t]
\centering
\includegraphics[width=0.95\textwidth]{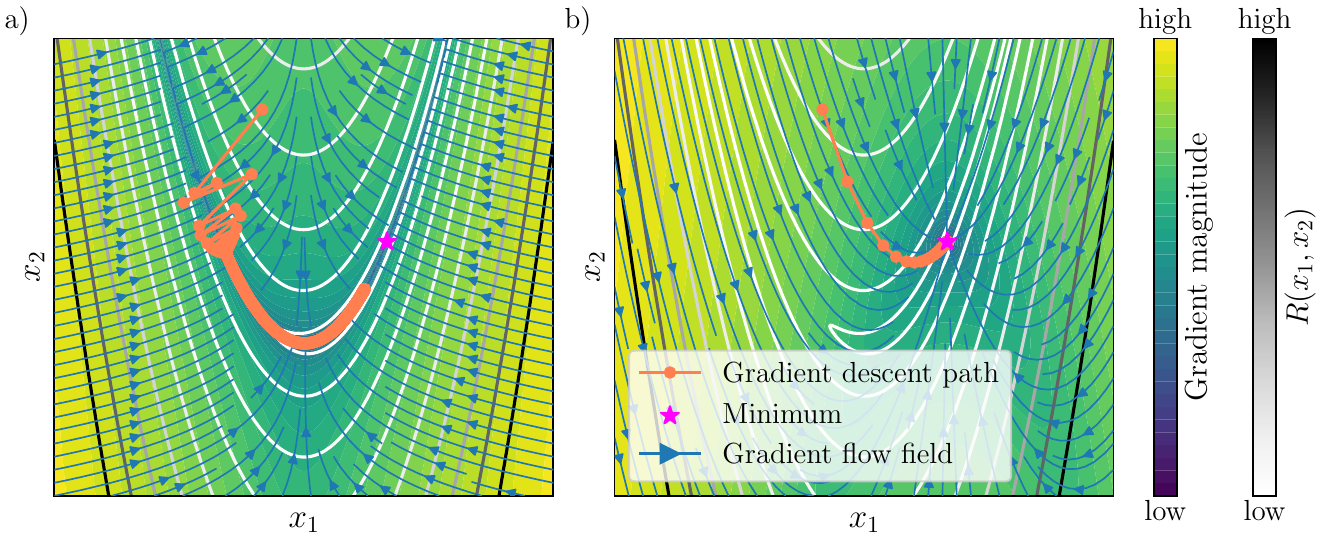}
\caption{
Here we compare gradient descent vs natural-gradient descent on the \textit{Rosenbrock landscape}. (a) Standard gradient descent $-\nabla R(x_1,x_2)$ produces a flow field with opposing directions that meet in a long valley. The opposing gradient directions produce a zig--zag trajectory as it repeatedly overshoots across the narrow curved valley of the Rosenbrock function (white contours), slowing convergence toward the minimum (magenta star).
(b) Natural-gradient descent $-S^{-1}\nabla R(x_1,x_2)$, with QGT $S$ as preconditioner, reshapes the flow to better align with the valley geometry, yielding a smoother path that follows the valley more directly. Background colors indicate the gradient magnitude (left color bar), while the grayscale bar encodes the objective value $R(x_1,x_2)$.
}
\label{fig:gdnat}
\end{figure}

\subsubsection{Ground state search}
\label{sec:gs_search}

Starting from the variational principle in Eq.~\eqref{eq:gs_energy}, ground-state search reduces to the numerical minimization of the variational energy with respect to the NQS parameters $\theta$. In practice, this is carried out with first-order gradient-based optimization. A simple and computationally cheap
choice is the \textit{Adam} optimizer, an adaptive stochastic optimization method widely used in machine learning \cite{kingmaAdamMethodStochastic2017}. Adam and related optimizers are robust, easy to implement, and perform well for moderately sized networks. In the NQS context they have been used successfully especially for autoregressive architectures such as RNNs \cite{hibat-allahRecurrentNeuralNetwork2025,lange2024,jreissaty2026}. However, autoregressive architectures are by no means limited to Adam-style first-order optimization, and can likewise be optimized with stochastic reconfiguration \cite{rigo25,malyshev2024}.

In fact, most state-of-the-art NQS ground-state results are obtained with \textit{stochastic reconfiguration} (SR) methods~\cite{sorella1998, becca2017}, which can be viewed as a stochastic realization of \textit{natural gradient descent} \cite{amari1998}. Physically, SR corresponds to imaginary-time evolution projected onto the variational manifold, in close analogy to the TDVP equation in Eq.~\eqref{eq:tdvp_equation}. This connection is reflected by the close similarity between the TDVP equation for the parameter time derivative $\dot{\theta}$ and the SR equation for the parameter update $\Delta\theta$,
\begin{equation}
\label{eq:sr_update}
\sum_j S_{ij}\,\Delta\theta_j = -\partial_i E(\theta).
\end{equation}
The parameters are then updated according to
\begin{equation}
\theta \leftarrow \theta + \eta\Delta\theta,
\end{equation}
with learning rate $\eta$.
The key idea is that the update is not chosen with respect to the Euclidean geometry of parameter space, corresponding to the identity metric $S_{ij}=\delta_{ij}$ used in ordinary gradient descent, but with respect to the geometry induced by the quantum state itself. This geometry is encoded in the \textit{quantum geometric tensor} (QGT),
\begin{equation}
\label{eq:SR}
S_{ij}
=
\mathrm{Re}\,
\bra{\partial_i \psi_\theta}\hat Q\ket{\partial_j\psi_\theta},
\qquad
\hat Q = 1 - \frac{\ket{\psi_\theta}\bra{\psi_\theta}}{\braket{\psi_\theta}{\psi_\theta}} \;,
\end{equation}
or, equivalently,
\begin{equation}
\label{eq:qgt_definition}
S_{ij}
=
\mathrm{Re}\,
\mathbb{E}_{\mathbf{s}\sim p_\theta}
\Big[
\big(\Gamma_i(\mathbf{s})-\mathbb{E}[\Gamma_i(\mathbf{s})]\big)^\dagger
\big(\Gamma_j(\mathbf{s})-\mathbb{E}[\Gamma_j(\mathbf{s})]\big)
\Big],
\end{equation}
where $\Gamma_i(\mathbf{s})=\partial_{\theta_i}\log\psi_\theta(\mathbf{s})$ and $p_\theta(\mathbf{s})\propto |\psi_\theta(\mathbf{s})|^2$. In this way, SR computes the projected imaginary-time update for which the residual of Eq.~\eqref{eq:SR} is orthogonal to the tangent space of the variational manifold $\text{span}\big(\lbrace\ket{\partial_i\psi_\theta}\rbrace\big)$, in direct analogy to the TDVP construction in Eq.~\eqref{eq:tdvp_equation}. As illustrated in Fig.~\ref{fig:gdnat}, this does not change the optimization landscape itself, but rescales updates so that broad weak-gradient directions do not slow down the optimization, thereby guiding the trajectory more directly toward the minimum.

The drawback of SR is its higher computational cost. In its standard form, one has to estimate and solve a linear system involving the $N_P\times N_P$ QGT, so the optimization bottleneck scales with the number of variational parameters. Since the Monte Carlo estimate of the QGT has rank at most $N_S$, this parameter-space formulation becomes prohibitive for modern deep architectures with $N_P\gg N_S$. Recent low-rank reformulations of SR, in particular \textit{minSR} \cite{chen2024,rende_2024}, exploit the structure of the stochastic QGT expression in Eq.~\eqref{eq:qgt_definition} and replace the $N_P\times N_P$ linear system by an $N_S\times N_S$ problem (or $2N_S\times2N_S$ for complex wave functions), thereby shifting the bottleneck from parameter count to sample count. Empirically, this reformulation is often already favorable for moderate sample numbers, and in many recent applications stable training was achieved with $N_S\sim10^3$, enabling deep architectures with up to $10^6$ parameters \cite{chen2024,viteritti2024,rende_2024,roth2025}.

In any formulation, the SR equation is typically ill-conditioned because the QGT contains very small eigenvalues arising from statistical noise or redundant parameter directions. In practice, this is mitigated by regularization, most simply through a diagonal shift $S\rightarrow S+\lambda \mathbb{I}$ with $\lambda>0$. Stability and convergence can be further improved by more refined regularization schemes. For example, SPRING (\textit{sub-sampled projected-increment natural gradient descent}) \cite{Goldshlager_2024} smooths the effective eigenvalue cutoff of the QGT and combines this with momentum, reducing sensitivity to Monte Carlo noise while accelerating convergence. Gu et al.~\cite{gu2025} later extended this idea with a dynamically adapted variant that performed particularly well for the Hubbard model, further improving the robustness and scalability of SR-based optimization.

\subsubsection{Convergence Metrics}

For the ground-state search with NQS, the most common figures of merit are the variational energy and the \textit{energy variance} $\mathbb{V}\big[H_{\rm loc}\big] = \big\langle (H_{\rm loc} - \langle H_{\rm loc} \rangle)^2 \big\rangle$.
By the variational principle, the variational energy provides an upper bound on the exact ground-state energy, while the energy variance vanishes for any exact eigenstate and therefore supplies an additional stringent consistency check. A small variance alone does not certify that the ground state has been found, but Ref.~\cite{Wu2024} empirically observed that the \textit{dimensionless variance} (V-score) is approximately proportional to the true energy error, namely the relative deviation between the exact and variational ground-state energies.
When MCMC is the sampler of choice, final accuracy checks should also be repeated with fresh, independently initialized chains to verify that they thermalize and reproduce the reported observables, otherwise apparently accurate energies can be artifacts of frozen or non-ergodic sampling rather than of an improved variational state~\cite{kamal2026comment}.

\subsubsection{Dynamics}
\label{sec:dynamics}

In time-dependent VMC (tVMC), all derivatives are estimated using VMC as described above. Unlike ground-state search, real-time dynamics does not steadily approach a fixed point. Instead, integration errors are propagated forward and can accumulate over time, so only very little noise can be tolerated in the MC estimate of the QGT and in the solution of the TDVP linear system Eq.~\eqref{eq:tdvp_equation}.
In practice, increasing the number of samples is generally not sufficient to obtain a well-conditioned full-rank QGT, making careful regularization essential \cite{schmitt2020quantum,hofmann_role_2022}.
Since the hard cutoff of a plain Moore-Penrose pseudo-inverse is incompatible with smooth solutions of the TDVP, established regularization schemes rely on generalized pseudo-inverses with smooth truncations of the eigenvalue spectrum \cite{schmitt2020quantum,schmittJVMCVersatilePerformant2022,medvidovic2023variational}.
As optimal values for the cutoff may vary with time, recent work introduced a scheme to adapt the regularization parameter dynamically while fixing the residual \cite{lagnese_neural_2026}.
In addition, Ref.~\cite{schmitt2020quantum} introduced an additional regularization to eliminate particularly noisy contributions from the TDVP equation.
Due to the non-linearity of the TDVP equation \eqref{eq:tdvp_equation}, it is generally imperative to employ integrators with adaptive time step. In many cases, unnecessarily small time steps can be avoided by employing the quantum geometric tensor to quantify the integration error in state space instead of parameter space \cite{schmitt2020quantum}.

By contrast, best practices for more recent alternatives such as p-tVMC \cite{sinibaldiUnbiasingTimedependentVariational2023}, t-NQS \cite{walleManybodyDynamicsExplicitly2024}, and neural Galerkin \cite{sinibaldiTimeDependentNeuralGalerkin2026} approaches are still much less established, so we restrict our discussion to TDVP-based dynamics. Among these newer methods, some practical guidance is beginning to emerge for p-tVMC, for instance in the recent work of Gravina \cite{gravinaNeuralProjectedQuantum2025a}.

\subsubsection{Sampling}
The NQS formalism avoids explicit summation over the full many-body Hilbert space by replacing it with averages over a set of configurations $\{\ket{\mathbf{s}_i}\}_{i=1}^{N_S}\sim p_\theta(\mathbf{s})$,
drawn from the Born distribution of the variational wave function. Observables and derived quantities are then obtained as MC averages of their respective estimators Eq.~\eqref{eq:mc_estimator} over these samples.
Importantly, any quantity whose estimator can be written in the form of the local observable estimator in Eq.~\eqref{eq:local_estimator} can be estimated efficiently, with the required number of samples increasing only polynomially with the desired precision \cite{Nest2009SimulatingQC}.

By contrast, quantities entering the ground-state search or TDVP, such as the energy gradient and the QGT, do not admit a formulation as local observable estimators.
These quantities can nevertheless be estimated by rewriting them as importance-sampled averages with respect to the Born distribution. For a quantity $X=\sum_{\mathbf s} x(\mathbf s)$, one may write
\begin{equation}
X=\sum_{\mathbf{s}:\,p_\theta(\mathbf{s})\neq 0} p_\theta(\mathbf{s})\,\frac{x(\mathbf{s})}{p_\theta(\mathbf{s})}\;,
\end{equation}
but the resulting estimator is only unbiased if $x(\mathbf{s})=0$ whenever $p_\theta(\mathbf{s})=0$. If $\psi_\theta(\mathbf{s})$ has roots on configurations where $x(\mathbf{s})\neq 0$, the Born distribution does not sample all relevant contributions and the estimator is therefore biased \cite{sinibaldiUnbiasingTimedependentVariational2023,medvidovicNeuralnetworkQuantumStates2024}. This issue is often negligible in ground-state calculations when one works in sectors where the variational state is non-zero everywhere, but it can become important in real-time evolution where nodes may proliferate during the dynamics \cite{sinibaldiUnbiasingTimedependentVariational2023}.
Recent work proposes to sidestep biased estimators by employing importance sampling techniques \cite{Krinitsin2026}.

Even when no formal bias is present, Monte Carlo estimators can have a very large variance. This applies both to local observables and to general importance sampled quantities. While the resulting estimates remain efficient in the complexity-theoretic sense, a large variance increases the practical sampling cost, because the statistical error decreases only as $N_S^{-1/2}$. A particularly unfavorable situation arises when the Born distribution is strongly peaked in the chosen basis. Then the MC samples explore only a small part of configuration space, while the estimator may receive appreciable contributions from a much broader region. In that case, many samples are required before averages converge reliably. For derivative-based quantities such as the QGT and the energy gradient this problem is often more severe, because their significant contributions in configuration space are typically broader than those sampled efficiently by the Born distribution itself \cite{malyshev2024}.

Equally important as the estimation itself is the generation of the underlying samples. For that purpose, Markov-chain Monte Carlo (MCMC) is most commonly employed. In the standard Metropolis--Hastings scheme, illustrated schematically in Fig.~\ref{fig:gpuacceleration}, one constructs a Markov chain by proposing updates of the current configuration $\mathbf{s}\to\mathbf{s}'$ and accepting or rejecting them probabilistically. In lattice models with discrete degrees of freedom, the most common proposals are local moves such as single-spin flips, local particle hops, or pair updates chosen to preserve conserved quantum numbers \cite{carleoSolvingQuantumManybody2017b,medvidovicNeuralnetworkQuantumStates2024}.
These updates are simple and architecture agnostic, but their efficiency depends strongly on how rapidly they explore the relevant part of configuration space. MCMC produces correlated samples; in practice, one must also monitor thermalization and autocorrelation times and discard an initial warm-up period before accumulating measurements.

An important alternative is available for autoregressive NQS \cite{hibat-allah2020}.
There, the joint probability of a configuration is factorized into conditionals, as indicated by Eq.~\eqref{eq:autoregressive_network}
\begin{equation}
 p_{\theta}(\mathbf{s}) = |\psi_{\theta}(s_1)|^2 |\psi_{\theta}(s_2|s_1)|^2 \dots |\psi_{\theta}(s_N|s_1, \dots, s_{N-1})|^2 =\prod_{i=1}^{N} p_\theta(s_i|s_{<i}),
\end{equation}
which allows one to generate configurations sequentially by ancestral sampling
\begin{equation}
    s_i \sim p_\theta(s_i | s_{<i})\, \qquad \text{with}~p_\theta(s_1) = p_\theta(s_1|const.) \;.
\end{equation}
The resulting samples are independent by construction and therefore avoid both thermalization and autocorrelation issues. In addition, autoregressive models permit sampling strategies such as \emph{Gumbel top-$K$ sampling} \cite{koolStochasticBeamsWhere2019}, which generate samples without repetition and can improve sample efficiency when the Born distribution is strongly concentrated, thereby alleviating the variance problem \cite{malyshev2024}.

\subsection{Complexity and GPU parallelization}
\label{sec:gpu}
\begin{figure}
    \centering
    \includegraphics[width=\linewidth]{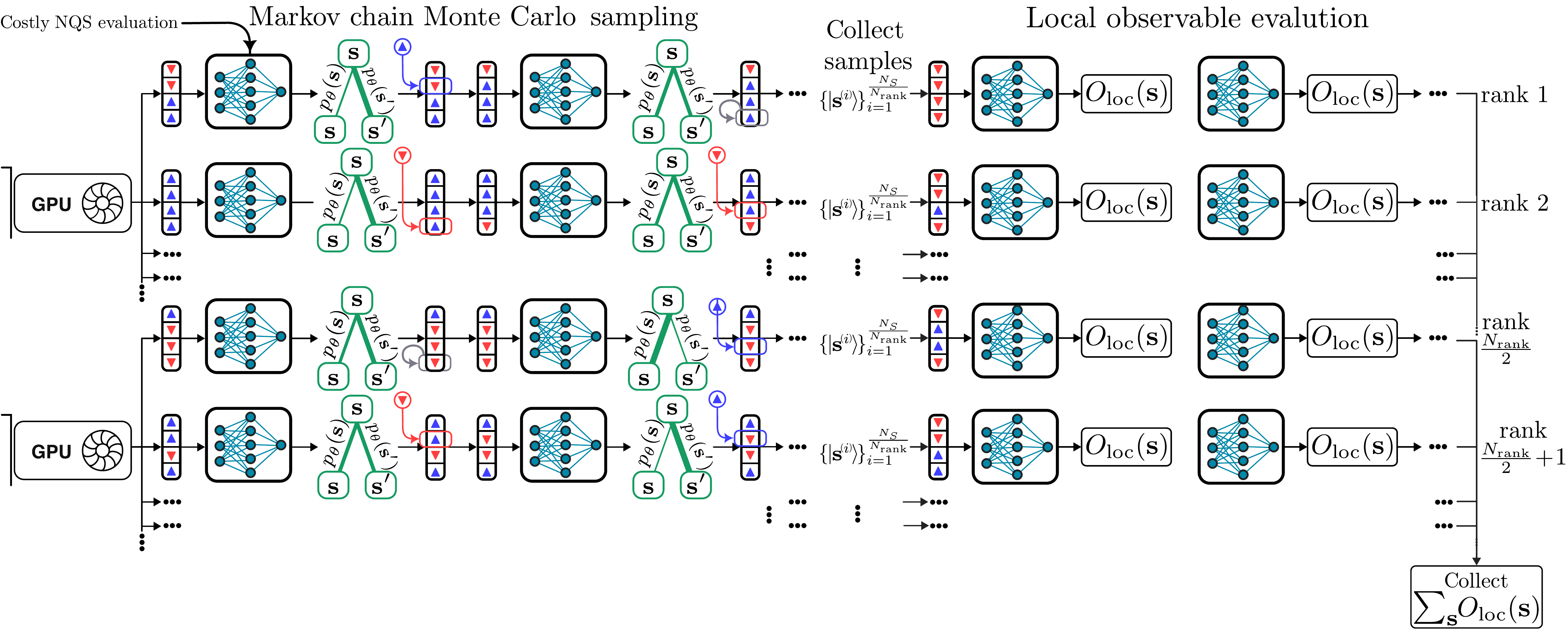}
    \caption{The figures show a schematic of the multi-GPU-accelerated workflow for NQS-based VMC or TDVP. The computationally dominant NQS evaluation is parallelized on GPUs during (left) Markov-chain Monte Carlo sampling and (right) local observable estimation. Multiple independent Markov chains are advanced concurrently by batching NQS evaluations on each device to better saturate the GPU. The Metropolis-Hastings accept/reject step is depicted as a probability-weighted binary decision tree, emphasizing the stochastic decision that turns a proposed configuration into the next configuration of the chain. The resulting sample pool $\lbrace \mathbf{s}_i\rbrace^{N_S/N_{\rm rank}}_{i = 1}$ is distributed across $N_{\rm rank}$ independent processes, which then independently evaluate local observables of the form $O_{\mathbf{s}\mathbf{s}'}\tfrac{\psi_\theta(\mathbf{s}')}{\psi_\theta(\mathbf{s})}$. Finally, per-process contributions are reduced to obtain the global Monte Carlo averages.}
    \label{fig:gpuacceleration}
\end{figure}

The algorithm underlying ground-state optimization with NQS, and likewise TDVP for real-time evolution, consists of three core stages: \emph{sampling}, \emph{estimation of observables and gradients}, and the \emph{optimizer step}. The total wall time of an algorithm iteration is therefore composed of sampling-and-estimation part and an optimization part,
\begin{equation}
    T_{\mathrm{ALG}}
    \sim
    T_{\mathrm{sample+obs}}
    +
    T_{\mathrm{opt}} \;,
\end{equation}
where $T_{\mathrm{sample+obs}} = n_{\mathrm{eval}}(N,N_S)\,T_{\mathrm{NQS}}(N)$ and $T_{\mathrm{NQS}}(N)$ denotes the cost of one \emph{batched} evaluation of the ansatz, that is the simultaneous evaluation of $\psi_\theta(\mathbf{s})$ for a whole set of configurations, and $n_{\mathrm{eval}}$ the number of such evaluations required per iteration.

The abstract decomposition above becomes more transparent by considering a typical example like a dense feed-forward ansatz (Fig.~\ref{fig:nqs_nutshell}) with width proportional to system size and depth $D$, for which one network evaluation scales as
\begin{equation}
    T_{\mathrm{NQS}}(N)\sim \mathcal{O}(DN^2).
\end{equation}
If sampling is performed with local Metropolis updates, then obtaining one effectively new sample requires roughly one sweep of $\mathcal{O}(N)$ proposed moves, so the sampling cost per sample is $\mathcal{O}(N\,T_{\mathrm{NQS}})$. For a local Hamiltonian on a $d$-dimensional lattice, the local-energy estimator couples a configuration to $\mathcal{O}(dN)$ connected configurations, so observable estimation likewise requires $\mathcal{O}(dN\,T_{\mathrm{NQS}})$ work per sample. The energy gradient can be obtained at essentially the same cost as the energy itself, since the required logarithmic derivatives are evaluated through automatic differentiation \cite{goodfellow2016}. Under local Metropolis sampling and without architecture-specific fast updates, this gives the illustrative estimate
\begin{equation}
    T_{\mathrm{sample+obs}}
    \sim
    \mathcal{O}\!\left(N_S D N^3\right),
\end{equation}
up to model-dependent prefactors and possible implementation-dependent speedups from caching, symmetry reduction, or architecture-specific fast updates \cite{chenLRUX2026}. At the same time, these stages are embarrassingly parallel, since each chain and each configuration can be processed independently, as sketched in Fig.~\ref{fig:gpuacceleration}. Distributing the work over $N_{\rm rank}$ parallel processes therefore reduces the wall time approximately to
\begin{equation}
    T_{\mathrm{sample+obs}}^{\rm wall}
    \sim
    \mathcal{O}\!\left(\frac{N_S D N^3}{N_{\rm rank}}\right),
\end{equation}
up to minor overhead. For MCMC, each chain incurs a warmup cost before measurements are accumulated. In optimization runs, however, chains are usually initialized from samples of the previous iteration, so once the parameters evolve smoothly only a short rethermalization is often needed, which is why parallelization becomes more effective as the sample count $N_S$ increases. Autoregressive samplers have the further advantage that they generate i.i.d.\ samples directly and therefore benefit from parallelization immediately.

The optimizer step can become a separate bottleneck when performing TDVP or ground-state search using (min)SR, since one must solve a linear system involving the quantum geometric tensor (QGT). In a parameter-space formulation with $N_P$ variational parameters, a dense QGT solve scales as $\mathcal{O}(N_P^3)$, which can become prohibitive for modern neural-network ans\"atze. Matrix-free iterative solvers avoid storing and factorizing the full QGT and instead reduce the problem to repeated QGT-vector products, substantially lowering the per-iteration cost. Their total runtime, however, still depends on the number of Krylov iterations and therefore on the conditioning of the problem and the regularization scheme \cite{carleoSolvingQuantumManybody2017b}. The minSR formulation exchanges the parameter dimension $N_P$ for the sample dimension $N_S$, which can be advantageous when $N_S \ll N_P$, but the same basic tradeoff between dense and matrix-free linear solvers remains. In practice, SR is therefore most effective when the linear solve is kept subdominant compared to sampling and observable estimation.

NQS are particularly well suited to hardware accelerators such as graphics processing units (GPUs), because the dominant workload consists of large batches of dense linear algebra and tensor contractions in the network evaluation and derivative computation. A single Markov chain or a single local-observable evaluation rarely saturates a modern accelerator, so many such tasks can be batched on one or multiple devices with little additional overhead, as illustrated in Fig.~\ref{fig:gpuacceleration}. This is precisely the execution model exploited by modern NQS packages such as \textsc{NetKet} \cite{vicentiniNetKet3Machine2022a}, \textsc{jVMC} \cite{schmittJVMCVersatilePerformant2022}, and \textsc{Quantax} \cite{Ao2026ChenAo}, which build on machine-learning frameworks like \textsc{JAX} \cite{jax2018github} to provide distributed execution and accelerator support with comparatively little problem-specific engineering.

\section{Future Directions}
\label{sec:futdir}
\begin{figure}[t]
    \centering
    \includegraphics[width=1\linewidth]{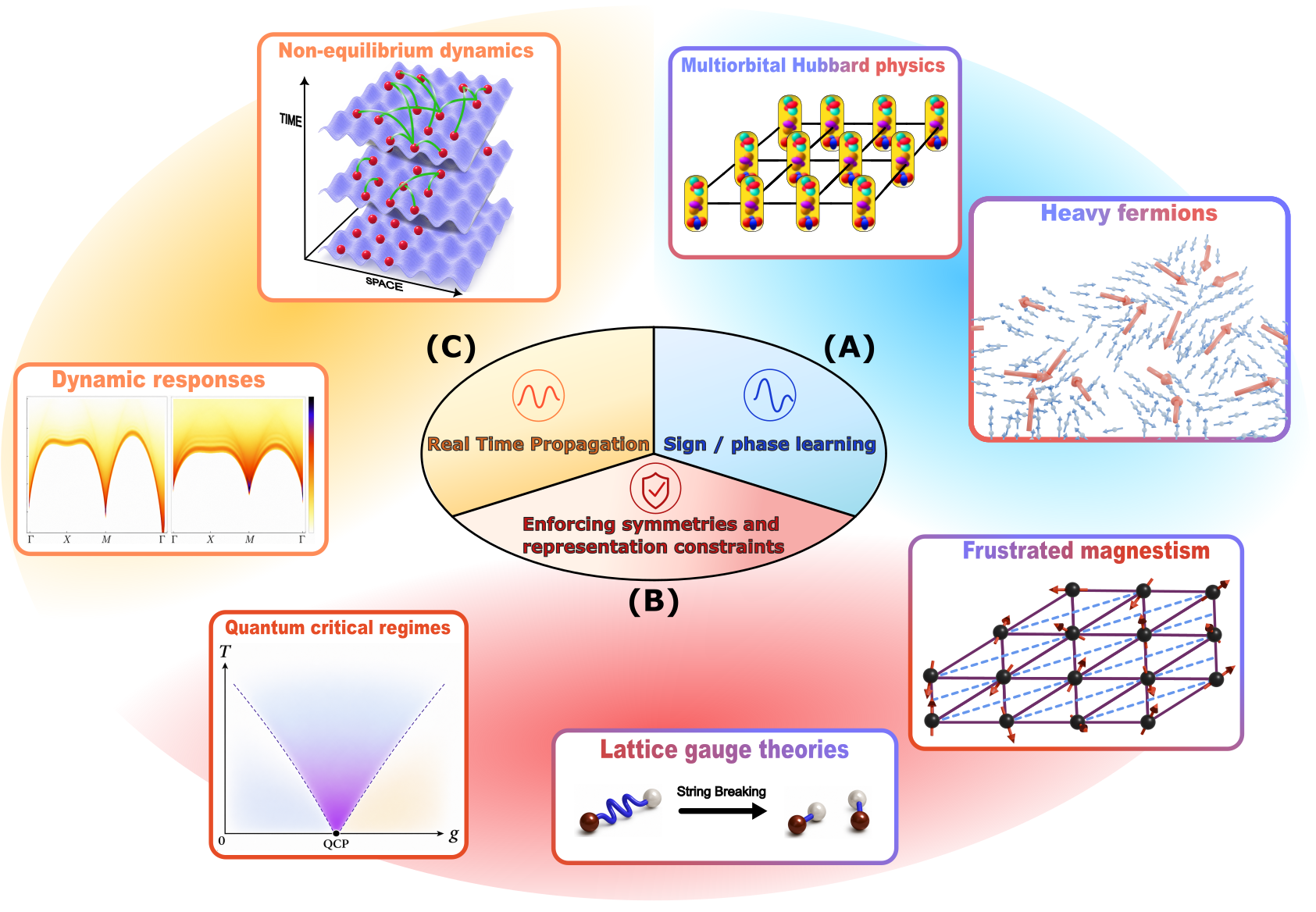}
    \caption{Signpost of future directions for NQS in condensed matter and related lattice problems. The central panel summarizes three current roadblocks that limit the application of NQS to key physical systems: (A) learning sign and phase structures, (B) enforcing symmetries and entanglement constraints, and (C) accessing real-time propagation and dynamical response. The surrounding panels illustrate problem classes that fall naturally within the scope suggested by versatile applicability and universal approximation capabilities, but whose treatment is still held back by one or more of these bottlenecks.}
    \label{fig:future_directions}
\end{figure}

As we have discussed in the previous sections, NQS is a versatile technique, able to address open questions at the forefront of condensed matter research.
The major advantages of NQS are:
\begin{enumerate}
    \item\textit{Versatile applicability across lattice geometries, entanglement structures, and irrespective of the QMC sign problem.}
    \item \textit{Universal approximation capabilities enabling controlled accuracy and non-perturbative treatment of strong correlations.}
\end{enumerate}
These two attributes define the realm where the most impactful future applications of NQS may be expected: Strongly correlated systems exhibiting intricate entanglement structures and a QMC sign problem.
Thereby, NQS could break new ground on a variety of current frontiers of condensed matter physics summarized in Fig.~\ref{fig:future_directions}.
However, the most demanding applications pose recurring challenges that will require further methodological advances, namely (A) learning sign and phase structures of the wave function, (B) enforcing symmetries and representation constraints, and (C) accessing dynamical response and non-equilibrium evolution.
In the following, we discuss these challenges and we outline possible future applications of NQS that would be enabled by overcoming them.

\subsection{(A) Learning sign and phase structures}

Sign and phase learning is a central bottleneck whenever the ground state has a non-trivial sign or phase structure, which is the case when the Hamiltonian, expressed in the sampling basis, has both negative and positive off-diagonal entries, a property known as \emph{non-stoquasticity}~\cite{bravyiComplexityStoquasticLocal2007}.
This offers a useful distinction when discussing sign problems of QMC versus NQS, since the non-stoquastic attractive Hubbard model can present a significant challenge for general NQS, but has a sign-problem-free auxiliary-field QMC formulation~\cite{becca2017,marvianComputationalComplexityCuring2019}.

Non-stoquasticity is a natural feature of fermionic Hamiltonians because of fermionic anticommutation relations, and it is also a common property of frustrated magnets.
In some important cases, however, the non-stoquasticity can be cured.
For example, in the square-lattice Heisenberg antiferromagnet, the Marshall sign rule fixes the ground-state sign structure~\cite{marshall1955}.
For free fermions, the antisymmetric signs are not a fundamental obstacle because the problem reduces to single-particle diagonalization.
The difficulty is that there is no generally efficient way to find such a cure: stoquastic local Hamiltonians form a special complexity class~\cite{bravyiComplexityStoquasticLocal2007}, and finding even restricted local basis rotations that cure a non-stoquastic Hamiltonian is NP-complete~\cite{marvianComputationalComplexityCuring2019}.
This leaves the vast majority of Hamiltonians in the seemingly untouchable class of non-stoquastic problems.

The situation is in practice more nuanced than that, as shown by a recent study of the isolated sign-learning problem, which demonstrates that frustrated sign structures can contain exploitable Boolean-Fourier structure and therefore need not behave like featureless worst-case learning problems~\cite{schurov2025}.
This nuance is further underlined by high-quality NQS ground states that have been obtained for non-stoquastic spin Hamiltonians close to simple proxy signs such as the Marshall sign rule~\cite{roth2023,chen2024,viteritti2025}.

The epitome of non-stoquastic systems currently treated with NQS is the class of fermionic Hamiltonians.
Here the standard strategy is to build antisymmetry into determinant, Pfaffian, or backflow-inspired layers~\cite{lou2019,moreno2022,lange2025,chen2025}.
These approaches represent the current state of the art for Hubbard-model benchmarks~\cite{gu2025,chen2025,viteritti2026variationalbiasresolvingintertwined,roth2025}, but they introduce a mean-field bias and a raw determinant or Pfaffian scaling cost of $\mathcal{O}(N_e^3)$, which low-rank-update techniques mitigate but do not remove~\cite{chenLRUX2026}.

This leaves two connected goals: (i) an unbiased treatment of sign and phase structures, and (ii) better scaling for controlled antisymmetric layers.
Progress on (ii) would already solve pressing problems that require large simulation sizes, like three-band cuprate models~\cite{emery1987,comp_persepective_HU_qin}, multilayer nickelates~\cite{li2019,nomura2022,nomuraStrongcouplingHighSuperconductivity2025}, and Moir\'e Hubbard models~\cite{wangmoire2023,luo2024simulatingmoirequantummatter}.
Progress on (i) would broaden access to elusive phases like quantum spin liquids~\cite{balents2010,lhuillier2011,knolle2019}, topological order~\cite{nayakNonAbelianAnyonsTopological2008}, and anyonic lattice models~\cite{Vieijra2020,luo2023gauge}.
Solving both goals is especially important near quantum criticality, where small energy differences between competing phases make careful finite-size scaling decisive~\cite{gegenwartQuantumCriticalityHeavyfermion2008a,kirchnerColloquiumHeavyelectronQuantum2020}.

\subsection{(B) Enforcing symmetries and representation constraints}
Historically, variational Monte Carlo wave functions were often built from physically motivated ans\"atze designed for a specific phase, such as Jastrow-correlated, Gutzwiller-projected, or mean-field states.
This makes their symmetry content and representation limits relatively transparent, but it also builds in strong assumptions about the phase being described~\cite{becca2017}.
That bias becomes limiting when the goal is to identify an unknown phase or resolve a delicate critical point.
A concrete example is the square-lattice N\'eel--VBS deconfined quantum critical point, where spin-rotation symmetry, lattice symmetries, and emergent critical symmetries all matter, while numerical scaling has required nonstandard interpretations such as two divergent length scales~\cite{wangDeconfinedQuantumCritical2017,shaoQuantumCriticalityTwo2016}.
NQS offer a possible route through this problem, since their large parameter spaces make systematic improvement with architecture and parameter count possible, but increasing the number of parameters does not fully address the issue.
For that, one must also know whether the architecture can represent the required symmetry sector and how its parameter count controls representational power.
This leaves two closely related challenges, namely symmetrizing NQS so that the Hilbert-space sector of interest is represented directly and finding laws that connect architecture and parameter count to representational power.

Recently, progress has been made on the latter in two significant ways. First, the entanglement that NQS can represent is now better understood through bounds such as $S_A=\mathcal{O}(k\log N)$ in the number of scalar nonlinearities $k$ and system size $N$~\cite{paul2026bound}. Second, transformer-NQS calculations on the square and triangular $J_1$-$J_2$ Heisenberg models found, within the accessible precision window, that the $V$-score decays approximately as a power law in training compute, with an exponent that decreases with frustration~\cite{rendeScalingLawsNeuralnetwork2026}.
Taken together, these results suggest that increasing model size and training compute can systematically improve NQS accuracy, while also indicating that frustration can make a given target accuracy more expensive.
The latter observation may be a symptom of an increasing number of near-degenerate low-energy states or symmetry-sector mixing~\cite{park2022a,westerhoutManybodyQuantumSign2023}.

Solving the problem of imposing symmetry constraints on NQS would make calculations more reliable when low-energy orders in different symmetry sectors compete~\cite{choo2018,nomura2021a}.
Currently, two complementary approaches exist for enforcing symmetries: projecting the ansatz into the right subspace with Eq.~\eqref{eq:symm}, or building the symmetry into an equivariant architecture such as a GCNN~\cite{cohen2016,roth2021}.
The architecture-based route can avoid explicit sums over all group elements and has been successful for lattice symmetries and, with more specialized equivariant constructions, also for non-Abelian gauge symmetries~\cite{roth2021,Romero2025,spriggs2025}.
Its accuracy and applicability, however, still depend on the expressive power of the chosen ansatz and on which representations or symmetry constraints are built in.
Projection-based approaches remain broadly applicable, including to non-Abelian or anyonic symmetry sectors, but they add a symmetry-averaging cost and can complicate optimization~\cite{choo2018,nomura2021a,Vieijra2020,Vieijra2021}.
Thus, whether symmetries are enforced projectively or by the underlying ansatz, the interaction between symmetry enforcement, expressivity, and optimization does not automatically improve results, but requires special care, as illustrated by the escalating symmetry projections used in~\cite{viteritti2024}.

Better handling of symmetry and representation constraints would make finite-size scaling more reliable near quantum criticality, where competing states can be very close in energy~\cite{gegenwartQuantumCriticalityHeavyfermion2008a,kirchnerColloquiumHeavyelectronQuantum2020}.
It would also sharpen calculations of quantum spin-liquid candidates and topologically ordered or anyonic phases, where the relevant sector is selected by symmetry, gauge, or fusion constraints as much as by energy~\cite{balents2010,knolle2019,wenColloquiumZooQuantumtopological2017,nayakNonAbelianAnyonsTopological2008}.
For lattice gauge theories, the same progress would let NQS work directly in the physical gauge-invariant sector for condensed-matter gauge descriptions and for Hamiltonian gauge dynamics in high-energy physics, including finite-density regimes where conventional Monte Carlo faces a sign problem~\cite{balents2010,brambilla2014,nagata2022}.

\subsection{(C) Accessing dynamical response and non-equilibrium evolution}
Investigating quantum many-body systems in non-equilibrium situations is a multifaceted research area addressing questions that range from the foundations of statistical physics to technological applications.
Theoretical studies are motivated by the development of increasingly more sophisticated experimental platforms for quantum simulation and quantum computing \cite{Bloch2008RMP,Georgescu2014RMP,Altman2021PRXQuantum} as well as ultrafast techniques to probe solid state systems \cite{Basov2017NatMater,deLaTorre2021RMP}.
Such experiments have, for example, revealed the possibility of light-induced superconductivity \cite{Fausti2011} or discrete time crystals \cite{Choi2017}, and in many cases, the experimental observations challenge our theoretical understanding.

Time evolution of quantum states combines different challenges discussed in the previous sections. In generic cases, the wave function rapidly develops non-trivial phase structures and the target state is typically surrounded by a continuum of mid-spectrum states.
Moreover, while the search for stationary low-energy states is often contractive and thereby forgiving deviations in individual optimization steps, forward-propagation in time requires high accuracy in every time step to avoid the accumulation of errors.
Simulations of linear response to weak perturbations as probed in various forms of spectroscopy, however, allow the exploitation of additional structure present within the low energy sector.
Promising NQS directions include real-time variational propagation, variational access to excited states, and direct reconstruction of dynamical correlators~\cite{mendes-santos2023,rigo25,medvidovic2025adiabatictransportneuralnetwork}.
Beyond ground-state response functions, scalable multi-state optimization, for example through orthogonalization, penalty terms, or enlarged-space formulations~\cite{valenti2022,Pfau_2024}, can turn NQS into a tool for computing experimentally relevant response functions in correlated electron materials, even for finite temperatures.

While various works have exploited the capabilities of NQS for novel insights into the non-equilibrium dynamics of quantum many-body systems beyond linear response \cite{schmittQuantumPhaseTransition2022,MendesSantos2024,Wiener2026,Naik2026,Naik20262}, severe bottlenecks have been reported in other cases \cite{King2025,Vovrosh2025}---the precise origin of which, however, remains unclear.
For further progress, it is therefore pivotal to develop a better methodological understanding and to address the identified limitations.
Sinibaldi et al. recently pointed out that conventional tVMC is formally ill-defined \cite{sinibaldiUnbiasingTimedependentVariational2023}, triggering various efforts to circumvent the problem by devising alternative ways of solving the time-dependent variational principle \cite{gravinaNeuralProjectedQuantum2025a, nys2024ab,Krinitsin2026}.
Developing a reliable toolbox with well-understood scope will be crucial for broader future applications.
Besides effective optimization, there is a substantial gap between the NQS network sizes used for time evolution (between $10^3$ and $10^5$ parameters) and the state of the art for ground-state problems (around $10^6$ parameters). As optimization cost is the limiting factor for parameter numbers in time evolution, scaling up network sizes is among the outstanding challenges---recall that enabling larger parameter counts by more efficient optimization was a key breakthrough for ground-state search \cite{chen2024}.
The common approach to assess accuracy of time-dependent simulations is testing agreement of observable quantities under varying network sizes and architectures and analyzing estimated deviations from the exact solutions (if available) in each time step.
These are, however, often difficult to interpret.
It would be desirable to develop an expressive figure of merit comparable across NQS architectures and model systems similar to the V-score for ground-state search \cite{Wu2024}.

Recent works have started exploring alternatives to the plain forward propagation of NQS in time.
The t-NQS \cite{walleManybodyDynamicsExplicitly2024} and neural Galerkin \cite{sinibaldiTimeDependentNeuralGalerkin2026} methods incorporate time as an explicit parameter of the wave function ansatz to solve Schrödinger's equation (or a Lindblad equation \cite{Vovk2026}) in the spirit of physics-informed neural networks \cite{Raissi2019}.
Besides possible routes to circumvent bottlenecks of conventional forward propagation, these methods may serve as a starting point for even more ambitious NQS modeling, such as foundation models that simultaneously solve the dynamics across a range of varying physical parameters \cite{Qi2026}.

Despite substantial room for methodological advances, NQS approaches are beginning to reveal new insights into the dynamical properties of quantum many-body systems.
Particularly promising directions connect to the strengths of NQS in targeting arbitrary lattice geometries and strong correlations---recall that dynamical situations are generally plagued by the MC sign problem.
The ability to obtain spectral functions in two- or three-dimensional systems could open new perspectives on the characteristics of unconventional equilibrium states of matter such as quantum spin liquids \cite{banerjee2017}, unconventional superconductors \cite{stewart2017} and quantum critical points \cite{kirchnerColloquiumHeavyelectronQuantum2020}.
Once pushed out of equilibrium, quantum many-body systems exhibit a rich phenomenology on transient timescales.
Light-induced order has been observed in various forms, but investigating the underlying microscopic mechanisms and developing a notion of transient non-equilibrium phases of matter remains an active area of research \cite{deLaTorre2021RMP}.
The ability to realize artificial quantum many-body systems in two or three dimensions poses numerous physical questions that advanced NQS techniques may help to shed light on, e.g., concerning phase ordering dynamics \cite{manovitz_quantum_2025}, complex relaxation dynamics of topological defects \cite{krinitsin_roughening_2025}, or time evolution subject to strong dynamical constraints \cite{Moudgalya2022,Adler2024}---including lattice gauge theories \cite{Aidelsburger2022,Meth2025}.
The ongoing developments in quantum computing and quantum simulation demand numerical reference simulations. While recent attempts have remained limited \cite{king_2025,Vovrosh2026,Haghshenas2026}, NQS provide a possible route to further raise the bar of classical simulation capabilities.
Besides benchmarking, NQS may serve as a tool to develop strategies for many-body quantum control and the stabilization of non-equilibrium states of matter in quantum simulators.

\section{Conclusion}

The state-of-the-art applications discussed in this perspective show that NQS have developed from a proof-of-principle variational ansatz into a competitive tool for selected strongly correlated lattice problems.
Their successes in frustrated magnets, doped Hubbard models, and non-equilibrium dynamics point to two useful organizing principles. First, NQS have \textit{\textbf{versatile applicability}} across lattice geometries, entanglement structures, and Hamiltonians for which conventional QMC suffers from a sign problem.
Second, they offer \textit{\textbf{universal approximation capabilities}}, so that the approximation error can in principle be reduced arbitrarily by enlarging and refining the variational family. In combination with the varying bias of different architectures, this property allows for the certification of obtained results via self-consistent convergence checks.

The same examples also make clear why NQS are not yet a black-box solver. The main roadblocks are learning non-trivial sign and phase structures, enforcing symmetries and physical constraints without sacrificing expressivity, and obtaining stable real-time evolution and dynamical response.
These are serious algorithmic challenges, but---to the best of today's knowledge---they differ in character from the conventional QMC sign problem or the entanglement bottleneck of tensor networks.
In QMC, the sign problem can make statistical errors grow exponentially for a fixed formulation, and in tensor networks, high-dimensional or highly entangled states can exceed the practical capacity of the representation. For NQS, the corresponding obstacles are more often questions of architecture, optimization, sampling, and diagnostics. They can still be prohibitive in practice, but current results suggest that they are not universal no-goes in the same sense.

This distinction is the main reason to continue working on the roadblocks summarized in Fig.~\ref{fig:future_directions}. Progress would let NQS shed light on long-standing problems such as spin-liquid diagnostics in frustrated magnets, competing orders in multi-orbital and Moir\'e correlated materials, heavy-fermion quantum criticality, lattice gauge dynamics, and light-induced or transient phases of matter. It would also make NQS valuable as classical verification tools for programmable quantum simulators and quantum computing platforms~\cite{Bloch2008RMP,Georgescu2014RMP,Altman2021PRXQuantum}, and as a route to theoretical predictions for ultrafast probes of quantum materials~\cite{Basov2017NatMater,deLaTorre2021RMP}.

\vspace{1cm}
\textit{Acknowledgments---} The authors acknowledge valuable input and contributions by Filippo Vicentini during early stages of writing.
This work was supported by the Deutsche Forschungsgemeinschaft (DFG, German Research Foundation) under Germany’s Excellence Strategy – Cluster of Excellence Matter and Light for Quantum Computing (ML4Q) EXC2004/2 – 390534769.
JBR and MS were supported via the Helmholtz Initiative and Networking Fund, grant no.~VH-NG-1711.

\pagebreak

\printbibliography
\end{document}